\documentclass[
superscriptaddress,
 amsmath,amssymb,
 aps,
prl,
twocolumn,
]{revtex4-2}

\newcommand{\thor}{$^{229}$Th}

\newcommand{\lisaf}{LiSrAlF$_6$}

\usepackage{graphicx}
\usepackage{gensymb}

\usepackage{upgreek}

\usepackage{float}  

\usepackage[dvipsnames]{xcolor}
\definecolor{ricky}{cmyk}{0, 0.7808, 0.4429, 0.1412}
\usepackage{xcolor}

\usepackage{braket}

\usepackage[version=3]{mhchem}

\usepackage{upgreek}

\begin{document}

\title{Theory of internal conversion of the \thor\ nuclear isomer in solid-state hosts}
\author{H. W. T. Morgan}
\affiliation{Department of Chemistry and Biochemistry, University of California, Los Angeles, Los Angeles, CA 90095, USA}
\affiliation{Department of Chemistry, University of Manchester, Oxford Road, Manchester M13 9PL, UK}
\author{H. B. Tran Tan}
\affiliation{Department of Physics, University of Nevada, Reno, Nevada 89557, USA}
\affiliation{Los Alamos National Laboratory, P.O. Box 1663, Los Alamos, New Mexico 87545, USA} 

\author{R. Elwell}
\affiliation{Department of Physics and Astronomy, University of California, Los Angeles, CA 90095, USA}

\author{A. N. Alexandrova}
\affiliation{Department of Chemistry and Biochemistry, University of California, Los Angeles, Los Angeles, CA 90095, USA}

\author{Eric R. Hudson}
\affiliation{Department of Physics and Astronomy, University of California, Los Angeles, CA 90095, USA}
\affiliation{Challenge Institute for Quantum Computation, University of California Los Angeles, Los Angeles, CA, USA}
\affiliation{Center for Quantum Science and Engineering, University of California Los Angeles, Los Angeles, CA, USA}
\author{Andrei Derevianko}
\email{andrei@unr.edu}
\affiliation{Department of Physics, University of Nevada, Reno, Nevada 89557, USA}
\date{\today} 

\begin{abstract}
Laser excitation of \thor{} nuclei in doped wide bandgap crystals has been demonstrated recently, opening the possibility of developing ultrastable solid-state clocks and sensitive searches for new physics. We develop a quantitative theory of the internal conversion (IC) of isomeric \thor{} in solid-state hosts and elucidate a crucial requirement in choosing the solid-state hosts  in nuclear clock applications. The IC of the isomer proceeds by resonantly exciting a valence band electron to a defect state, accompanied by multi-phonon emission. We demonstrate that, if the process is energetically allowed, it generally quenches the isomer on a millisecond timescales, much faster than the isomer’s radiative lifetime, despite \thor{} being in the +4 charge state in the valence band.
\end{abstract}

\maketitle

The laser-accessible 8.4 eV isomeric transition in \thor\ nucleus holds an intriguing promise for developing quantum technologies and fundamental physics searches based on coherent manipulation of {\em nuclear} degrees of freedom. 
In particular, a crystal doped with a macroscopic number of \thor\ nuclei enables realizing a portable frequency standard with an unprecedented degree of stability - a solid-state nuclear clock~\cite{Rellergert2010}.  
Observations of direct laser excitation of \thor{} in three distinct solid-state hosts have been reported recently~\cite{Tiedau2024,Elwell2024,zhang2024thf,Zhang2024-Th229Comb}.

\begin{figure}
    \centering
    \includegraphics[width=0.4\textwidth]{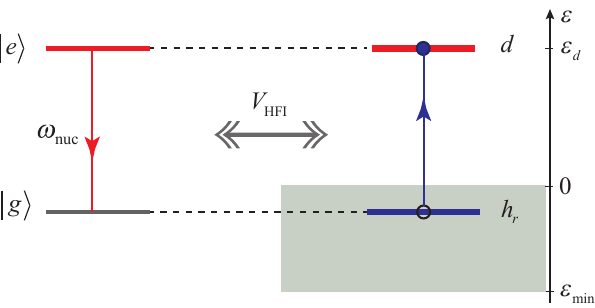}
    \caption{During the internal conversion in a crystal, a \thor{} nucleus relaxes non-radiatively by resonantly transferring its $\omega_\mathrm{nuc} \approx 8.4 \,\mathrm{eV}$ energy to a particle-hole ($d-h_r$) electronic excitation. The hole state $h_r$ lies in the valence band, while the defect state $d$ lies inside the bandgap.
    The excitation transfer is mediated by $V_\mathrm{HFI}$, the hyperfine coupling   between the nuclear and electronic degrees of freedom. The valence band is shown as a gray box, with the valence band maximum at $\varepsilon=0$ and the minimum at $\varepsilon_\mathrm{min}$.}
    \label{fig:IC-sketch}
\end{figure}

There is presently great need for a quantitative theoretical understanding of processes in these novel systems so that they may be optimized.
The challenge lies in bridging concepts and techniques from distinct sub-fields: materials science, quantum chemistry, condensed matter physics, nuclear physics, and atomic and optical physics. 
In particular, one of the critically important decay channels in nuclei in a chemical environment is the internal conversion (IC) channel,  see Fig.~\ref{fig:IC-sketch}.  
During IC, the nucleus relaxes non-radiatively by transferring its energy to the environment. 
As such, IC is practically important, as it competes with the radiative nuclear decay channel and can greatly affect clock performance.
While IC in a free \thor{}  atom and its ions has been understood quantitatively~\cite{Karpeshin2007,Tkalya2015,Bilous2017,Bilous2018}, the IC theory of  \thor{} ions in a crystal environment has been lacking~\cite{TkaSchJee2015}. 
This work aims to fill this gap. 

There are contradicting qualitative arguments in the literature regarding the relative importance of the IC process in solid-state hosts. 
For example, Ref.~\cite{Tiedau2024} does ``not  expect a significant contribution of bound internal conversion to the decay rate.''  
Meanwhile, Ref.~\cite{Elwell2024} postulates a quenching mechanism due to electronic defect states introduced into the crystal in the \thor{} doping process; likewise the results of Ref.~\cite{Pineda2024} may be interpreted to support the presence of non-radiative decay. 
Here, we analyze the IC process theoretically and show that, if energetically allowed, the IC process in \thor{}-doped crystals occurs on a millisecond timescale, much faster than the $\sim 1000 \, \mathrm{s}$ spontaneous radiative decay.  
We develop a general framework by combining the widely used density functional theory (DFT) with the {\em ab initio} relativistic treatment of electron interactions with the \thor{} nucleus, capturing the important ingredients of IC in \thor{}-doped solid-state hosts.

In solid-state \thor\ experiments, suitable host crystals are insulators, with band-gaps larger than the nuclear transition frequency $\omega_{\mathrm{nuc}} \approx 8.4 \,\mathrm{eV}$  so that the crystals are transparent to VUV radiation~\cite{Rellergert2010} -- \ce{ThF4} in Ref.~\cite{zhang2024thf}, \lisaf{} in Ref.~\cite{Elwell2024}, and CaF$_2$ in \cite{Tiedau2024,Zhang2024-Th229Comb}.
Doping \thor{} into \lisaf{} and \ce{CaF2} generically creates spatially-localized electronic defect states with typical  energies inside the insulator gap, as shown in Fig.~\ref{fig:IC-sketch}. 

The electronic energies and wavefunctions (including those of defect states) in solid state materials may be obtained by solving the eigenvalue equation
\begin{equation}\label{eq:Schrodinger_eqn}
h_{\mathrm{el}}\left(  \mathbf{r}\right)  \phi_{i}\left(  \boldsymbol{r}%
\right)  =\varepsilon_{i}\phi_{i}\left(  \boldsymbol{r}\right)  ,
\end{equation}
where $h_{\mathrm{el}}\left(  \mathbf{r}\right)  $ is a suitably chosen self-consistent field
electronic Hamiltonian and $\phi_{i}$ are single-electron orbitals. In materials science, density functional theory (DFT) is widely used to handle Eq.~\eqref{eq:Schrodinger_eqn}.
Kohn-Sham DFT gives orbitals $\phi_{i}$ for the valence band and the defect at a {fixed lattice and doping} geometry, and all orbitals are orthogonalized during the calculation.
For more details on our use of DFT in \thor:\lisaf{} see the supplementary information (SI).

The valence band is fully occupied and in the second quantization it can be represented by the quasi-vacuum state $|\Omega\rangle = \left( \prod_{\varepsilon_i <0} a^\dagger_i\right) |0\rangle$, where we placed the zero of energies $\varepsilon_{i}$ at 
the valence band maximum. 
The electronic
Hamiltonian may be written as
$
H_{\mathrm{el}}=\sum_{k}h_{\mathrm{el}}\left(  \mathbf{r}_{k}\right)
=\sum_{i}\varepsilon_{i}:a_{i}^{\dagger}a_{i}:
$.
Here $a^{\dagger}$ and $a$ are the fermionic creation and annihilation operators and  $:\cdots:$ stands for the conventional normal ordering of operator products~\cite{FetWal71} with respect to the quasi-vacuum state $|\Omega\rangle$. The summation with $i$ extends over the entire single-electron spectrum $\{\varepsilon_{i}\}$.  
Of particular interest for this work are states where a valence electron is excited into a defect state above the Fermi level, leaving behind a ``hole''.  A particle-hole excitation from the valence band to the defect state $d$ reads $a_{d}^{\dagger}a_{h}|\Omega\rangle$. 
The energy of this particle-hole excitation is $\varepsilon_{d}-\varepsilon_{h}$.

Here, the nuclear subsystem is modeled as two distinct energy levels: the ground state $|g\rangle$ with nuclear spin $I_g = 5/2$ and the excited (isomeric) state $|e\rangle$ with $I_e = 3/2$, separated by the energy gap $\omega_{\mathrm{nuc}}$. 
Fixing the energy of the nuclear ground state at zero, the unperturbed nuclear
Hamiltonian reduces to $H_{\mathrm{nuc}}=\omega_{\mathrm{nuc}}|e\rangle\langle e|$, with an implicit summation over nuclear magnetic sublevels, $m_I$. 
Unless specified otherwise, atomic units, $|e|=\hbar=m_e \equiv 1$, are used throughout.

The Hilbert space of the compound electron-nuclear system is spanned by the tensor product of the nuclear and electronic states 
with the absolute ground state of the compound system, $H_0 = H_{\mathrm{el}}+ H_{\mathrm{nuc}}$, written as
$|\Psi_{0}^{g}\rangle=|\Omega\rangle|g\rangle.$
The nuclear and electronic degrees of freedom are coupled via the hyperfine interaction (HFI) $V_{\mathrm{HFI}}$ and the total Hamiltonian of the compound
system reads
$H=H_{\mathrm{el}}+H_{\mathrm{nuc}}+V_{\mathrm{HFI}}$,
with
\begin{equation}
V_{\mathrm{HFI}}=\sum_{ij}\sum_{n^{\prime}n}V_{ij}^{n^{\prime}n}%
:a_{i}^{\dagger}a_{j}:|n^{\prime}\rangle\langle n| \,,
\end{equation}
where indices $i$ and $j$ range over the single-electron spectrum~\eqref{eq:Schrodinger_eqn} and $n$ and $n^{\prime}$ run over the nuclear states $|e\rangle$ and $|g\rangle$. 
Phononic degrees of freedom are suppressed in this expression, but included in the calculations that follow.

The main task for analyzing the impact of IC decay is to understand the lifetime of the nuclear state in a crystalline environment. 
Thus, our aim is to calculate the lifetime of the state, $|\Psi_{0}^{e}\rangle=|\Omega\rangle|e\rangle$, with the nucleus in the isomeric state $|e\rangle$ and all the electrons occupying the valence band, which has energy $E_{0}^{e}=\omega_{\mathrm{nuc}}$. 
Such a state is embedded into the continuum of particle-hole states%
\begin{equation}
|\Psi_{dh}^{g}\rangle=a_{d}^{\dagger}a_{h}|\Omega\rangle|g\rangle \,, \, E_{dh}^{g}=\varepsilon_{d}-\varepsilon_{h} 
\end{equation}
with the nucleus in its ground state, an electron in the defect state, and a hole in the valence band. 
Given the valence band is a continuum, typically one of these states, denoted as $|\Psi_{dh_r}^{g}\rangle$
where $h_{r}$ is the resonant hole (see Fig.~\ref{fig:IC-sketch}) has an energy of 
\begin{equation}\label{eq:res_cond}
E_{dh_r}^{g}=\varepsilon_{d}-\varepsilon_{h_{r}} = \omega_{\mathrm{nuc}} \,,
\end{equation}
and is therefore degenerate with $\ket{\Psi_0^e}$.

In the illustrative model of Fig.~\ref{fig:IC-sketch}, the discrete spectrum contains a single defect state. 
Since the zero of electronic energy is defined at the valence band maximum (i.e., the Fermi level), the energy of the hole must satisfy $\varepsilon_{h_{r}}\leq 0$.
As a result, the resonance condition~\eqref{eq:res_cond} is only met if the defect lies at or below the nuclear excitation on the energy diagram, leading to the requirement:
\begin{equation}
\varepsilon_{d}\le \omega_{\mathrm{nuc}}.\label{Eq:Energy-condition}
\end{equation}
In addition, the energy of the resonant hole must lie within the valence band, implying 
\begin{equation}
\varepsilon_{d} -\varepsilon_\mathrm{min}  \geq \omega_{\mathrm{nuc}},\label{Eq:Energy-condition-min}
\end{equation}
where $\varepsilon_\mathrm{min}$ is the valence band minimum, see Fig.~\ref{fig:IC-sketch}.

This model maps onto a textbook problem~\cite{CohTan98}, where a discrete state $|\Psi_{0}^{e}\rangle$ is embedded into the $|\Psi_{dh}^{g}\rangle$ continuum.  
These are coupled by a time-independent perturbation $V_{\mathrm{HFI}}$. 
As a result, the $|\Psi_{0}^{e}\rangle$ state decays into the continuum, relaxing the nucleus, and generating a particle-hole pair. 
This is an internal conversion (IC) mechanism in the solid-state hosts. 
The rate of this decay is given by Fermi's golden rule~\cite{cohen1998atom}%
\begin{align}\label{eq:simple rate}
\Gamma_{\mathrm{IC}}  & =\frac{2\pi}{\hbar} \rho\left(  \varepsilon_{h_{r}%
}\right)  \left\vert \langle\Psi_{dh_{r}}^{g}|V_{\mathrm{HFI}}|\Psi_{0}%
^{e}\rangle\right\vert ^{2}\nonumber\\
&= \frac{2\pi}{\hbar}\rho\left(  \varepsilon_{h_{r}} = \varepsilon_{d} - \omega_\mathrm{nuc} \right)  \left\vert
V_{dh_{r}}^{ge}\right\vert ^{2}\,,
\end{align}
with $\rho\left(  \varepsilon_{h_{r}} \right)$ being the electronic density of states (DOS) at the resonant hole energy. {The time-reversed analog of the IC process is the Nuclear Excitation by Electron Transition (NEET) process; it discussed in the SI.}

The decay rate~\eqref{eq:simple rate} assumes a single defect state. For a manifold $\left\{  d_{i}\right\}  $ of defect states, the rate is summed over all available decay channels. 
Summing over the nuclear magnetic quantum
numbers $m_{g}$  of the final state and averaging over magnetic
quantum numbers $m_{e}$ of the initial nuclear state, the rate becomes:
\begin{align}\label{eq:rate_Fermi}
\Gamma_{\mathrm{IC}}&=\frac{2\pi}{2I_{e}+1}\sum_{d_{i}%
}\sum_{h_{r}}\rho\left(  \varepsilon_{h_{r}}=\varepsilon_{d_{i}}%
-\omega_{\mathrm{nuc}}\right) \nonumber\\
&\times\sum_{m_{e}m_{g}}\left\vert V_{d_{i}h_{r}%
}^{gm_{g}em_{e}}\right\vert ^{2}\,.
\end{align}
Here, the energy selection rules  (\ref{Eq:Energy-condition},\ref{Eq:Energy-condition-min}) are included implicitly, as the DOS vanishes if these conditions are violated.

\begin{figure}
    \centering
    \includegraphics[width=0.45\textwidth]{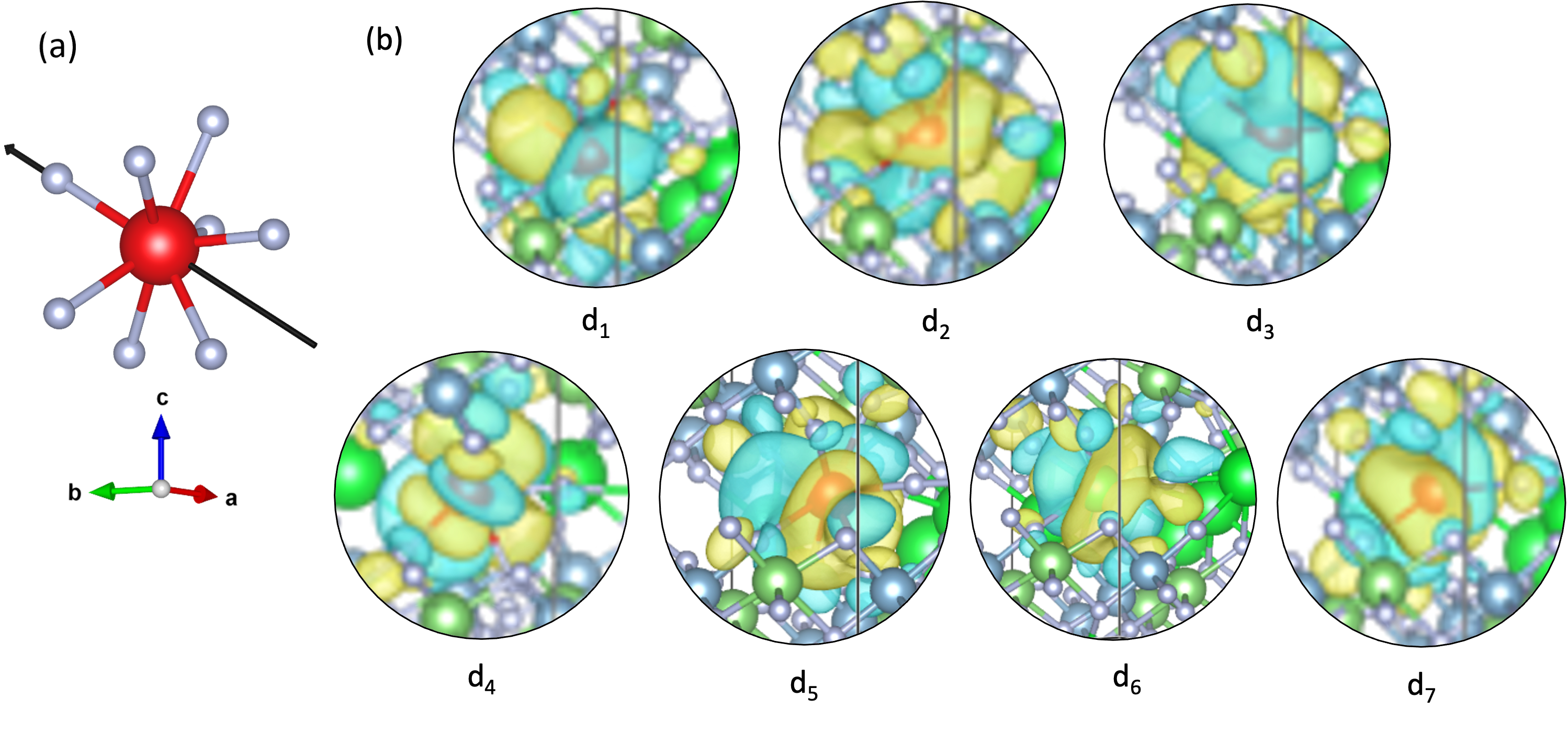}
    \caption{(a) \ce{ThF8} cluster from the optimized structure of Th:\lisaf{} annotated with the vector $(0.01,0.135,0.065)$ in black in fractional coordinates, which is a vector from the Th atom to one of the nearest-neighbor F atoms.
 (b) Real-space wavefunctions of the Th defect states evaluated at the $\Gamma$ point. {These closely resemble $5f$ electronic orbitals.}
 Th atoms are shown in red, Li in dark green, Sr in bright green, Al in blue, and F in grey. Blue and yellow isosurfaces represent the real part of the wavefunction - in all cases the imaginary part appears identical.}
    \label{fig:Th-defects-5f-orbitals}
\end{figure}

For concreteness, we evaluate the IC rates for Th-doped \lisaf{} and \ce{CaF2}. The technique we develop below applies to a broad class of Th-doped VUV transparent solid-state hosts.
The electronic structure of Th-doped \lisaf\ is computed within DFT using the Vienna Ab initio Simulation Package (VASP)~\cite{VASP}. 
Details are provided in SI and Ref.~\cite{Elwell2024}. 
We consider a doping geometry with one Th atom substituted into a Sr site and two additional interstitial F atoms (for charge compensation). 
Such a doping geometry yields the lowest energy among various possible site substitutions. 
These calculations find that the lowest-energy unoccupied orbitals are a closely-spaced manifold of seven states $\left\{d_{i}\right\}$ resembling the $5f$ orbitals $\psi_{5fm}\left(  \boldsymbol{r}\right)$ of thorium (here $m$ denotes the magnetic quantum number).
Fig.~\ref{fig:Th-defects-5f-orbitals} shows real-space representations of these orbitals.
The Th $6d$ orbitals appear above the $5f$ orbitals.
In the ground electronic state these orbitals are empty, so Th is in the +4 oxidation state. 
In the excited electronic state $|\Psi_{dh}\rangle$ an electron is transferred from F into the Th $5f$ manifold, putting Th into the $+3$ oxidation state.
This is evidenced by the 5$f$-like unoccupied states in shown in Fig. \ref{fig:Th-defects-5f-orbitals} and the predominant F 2$p$ character of the valence band shown in the PDOS in Fig. S1. 

\begin{figure}
    \centering
    \includegraphics[width=0.5\textwidth]{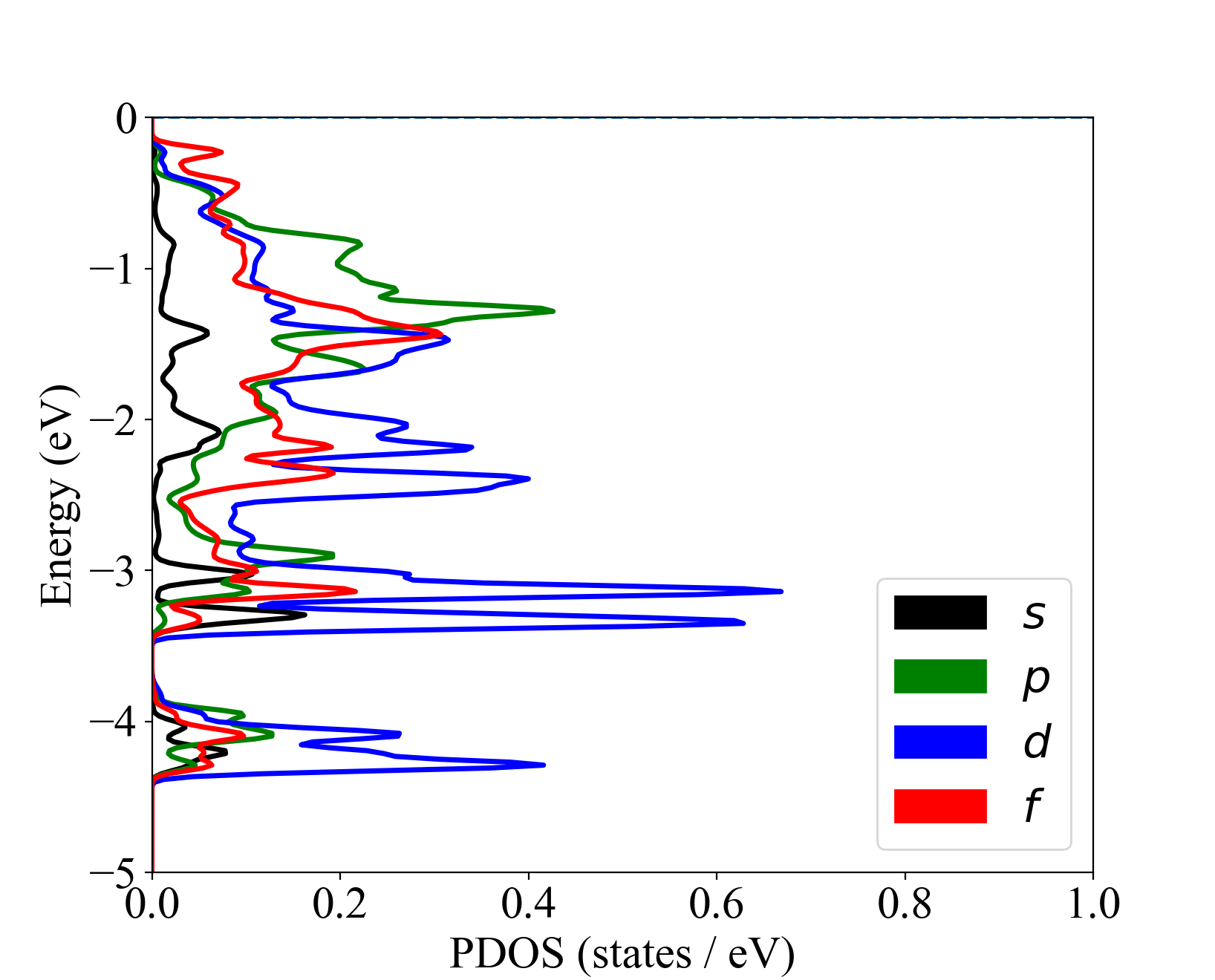}
    \caption{Thorium projected density of states in the valence band for Th:\ce{LiSrAlF6}. 
    }
    \label{fig:valence band Th PDOS}
\end{figure}

An IC-induced excitation of the valence band electron to the defect state is accompanied by a charge transfer from F to the Th ion. 
This leads to electrostatic distortion of the lattice, see SI.
In a molecular picture of the defect, the vibrational state of the \ce{ThF_n} ``cluster'' is significantly excited after nuclear relaxation, leading to multi-phonon relaxation. 
{Our computed lattice relaxation energy of the Th$^{3+}$-like electronic defect states is $E_R\sim 0.5 \,\mathrm{eV}$, while the Th-localized normal mode frequencies are $\sim 70\,\mathrm{meV}$, estimated based on the Th-F stretching  mode frequencies in the ThF$_n$ molecules~\cite{Andrews2013}. Lattice relaxation requires emitting at least $\approx 7$ phonons.}
The evaluation of the IC rate must therefore include the phononic degree of freedom and summing over the final phonon states. We find that the rate is modified by a factor of order unity. 
Therefore, 
the IC rate~\eqref{eq:rate_Fermi} is valid without including phononic effects at low temperatures, but with the defect energies shifted by the relaxation energy $\varepsilon_{d_i} \rightarrow \varepsilon_{d_i}  - E_R$. 
In the following, this shift is suppressed for brevity.

We repeated the DFT calculations for Th:\ce{CaF2}, 
using the geometry with two interstitial fluorides at 90\degree that was previously shown to be the lowest energy thorium defect~\cite{RN611}.
PDOS plots for Th and F in Figs. S2-4 show that the electronic structure of thorium is qualitatively similar in Th:\ce{CaF2} and Th:\ce{LiSrAlF6} and that both will undergo F-Th electron transfer excitations. In Th:\ce{CaF2} we also find the Th $5f$-like orbitals to be the lowest unoccupied orbitals, with Th in the $+4$ oxidation state. This motivates our treatment of the electronic defect states as resembling the Th$^{3+}$ $5f$ orbitals $\psi_{5fm}(\boldsymbol{r})$. 
Indeed, the $4f$ orbitals are included in the frozen core of thorium, while the energies of the $6f$, $7f$, $\ldots$ orbits are too high to contribute significantly to the valence band.

The HFI matrix element in Eq.~\eqref{eq:rate_Fermi} may be decomposed into a scalar product of the electronic and nuclear tensors of various ranks: 
\begin{equation}\label{eq:V_HFI}
V_{dh_{r}}^{ge}=\langle g|\boldsymbol{\mu}|e\rangle\cdot\langle
d|\boldsymbol{T}|h_{r}\rangle\,,
\end{equation}
where $\boldsymbol{\mu}$ is the nuclear M1 operator and $\boldsymbol{T}$ is an electronic  operator {(see SI)}. Inclusion of the higher-rank electric-quadrupole HFI would only increase the IC rate, strengthening the conclusions of our work.

Since the HFI matrix element is accumulated near the nucleus, we are interested in components of the valence band hole state centered on the thorium impurity. 
As the defect state is predominantly of the $5f$ character, we are interested in the $f$ content of the hole state due to the HFI selection rules ($\boldsymbol{T}$ is a rank 1 parity-even tensor). 
To this end, one may carry out a partial wave expansion,
\begin{equation}
\phi_{h}\left(  \boldsymbol{r} \right)  = \sum_{m_h} c^h_{5fm_h} \psi_{5fm_h}\left(  \boldsymbol{r}\right) + \ldots \,,
\label{Eq:expansion}
\end{equation}
where ellipses denote remaining contributions to the valence orbital $\phi_{h}$. 
In general, the $p$ admixtures in \eqref{Eq:expansion} can also contribute to the IC rate, but the HFI matrix elements between the $p$ and $f$ orbitals vanish non-relativistically and thereby contribute at a smaller level.  

In principle, the expansion coefficients may be found using
 $   c_{5fm}^h = \int d^3 r \,
    \psi_{5fm}^*\left(  \boldsymbol{r}\right) \phi_{h}\left(  \boldsymbol{r} \right)
 $.   
This, however, requires detailed knowledge of the valence band wavefunctions at distances comparable to the atomic size. 
At these sub-Bohr-radius scales, VASP utilizes pseudo-potentials 
that do not capture the correct behavior of atomic orbitals. 
Instead, we notice
that the projected density of states (PDOS) in the valence band for the $f$-orbital character reads 
\begin{equation}
    \rho_{f}(\varepsilon) \approx \rho_{5f}(\varepsilon) = 2 \sum_h \sum_{m_h} |c^h_{5fm_h}|^2 \delta( \varepsilon -\varepsilon_h)\,,
\end{equation}
which includes the electron spin degeneracy. 
This PDOS is plotted in Fig.~\ref{fig:valence band Th PDOS} (red curve). 
Since the DOS $\rho(\varepsilon) = 2 \sum_h \delta( \varepsilon -\varepsilon_h)$, we may relate the DOS and PDOS as
\begin{equation}
 \rho_{5f}(\varepsilon_{h}) \approx \rho(\varepsilon_{h}) \sum_{m_h} |c^{h}_{5fm_h}|^2 = 7 \rho(\varepsilon_{h}) 
 \langle |c^{h}_{5f}|^2 \rangle \label{Eq:PDOS-c2DOS}
 \,
\end{equation}
in the assumption of slowly varying expansion coefficients over energy and the introduction
of an average value $\langle |c^{h}_{n\ell}|^2 \rangle  =  \sum_{m} |c^{h}_{n\ell m}|^2/(2\ell +1) $.

With the hole-state expansion~\eqref{Eq:expansion} and the HFI matrix element~\eqref{eq:V_HFI}, the IC rate simplifies to
\begin{align}\label{eq:IC_rate_NR}
&\Gamma_{\mathrm{IC}}=\frac{2\pi}{2I_{e}+1}
\rho\left(  \varepsilon_{h_{r}}\right) 
\sum_{m_h }|c_{5fm_h}|^2\\
&\times\sum_{m_d m_{e} m_{g} m_s m_s'} 
\left\vert 
\langle gm_g |\boldsymbol{\mu}|em_e \rangle\cdot\langle
5f m_d m_s |\boldsymbol{T}|5f m_h m_s' \rangle
\right\vert ^{2} \,, \nonumber
\end{align}
where the $h_r$ superscript in the expansion coefficients has been dropped.

{Since the HFI matrix elements are accumulated near the nucleus and Th electrons in this region ($\alpha Z \approx 0.65$ for $Z=90$ of Th) are relativistic}, we derived the relativistic generalization of the expression~\eqref{eq:IC_rate_NR}, where we distinguish between two fine-structure components of the $5f_j$ state, $j=5/2$ and $j=7/2$, as 
\begin{align}
\label{eq:IC_rate_Relativistic}
\Gamma_{\mathrm{IC}} &=\frac{2\pi \rho\left(  \varepsilon_{h_{r}}\right)}{3(2I_{e}+1)} 
  \left\vert 
\langle g||\mu||e\rangle\right\vert ^{2}\nonumber\\
&\times  \sum_{j_dj_h} \left\vert \langle 5f_{j_d}||T||5f_{j_h}\rangle \right\vert ^{2} \langle |c_{5f_{j_h}}|^2 \rangle\,.
\end{align}
Here, the Wigner-Eckart theorem and the averages
$    \langle |c_{5f_{j_h}}|^2 \rangle \equiv  \sum_{m_h} |c_{5f_{j_h} m_h}|^2/ (2j_h +1)$
have been used to simplify the expression.
Since VASP is a non-relativistic code, we relate this average to the non-relativistic expansion coefficient $c_{n \ell m_l}$, c.f. Eq.~\eqref{Eq:expansion}, using the angular momentum algebra
\begin{equation}
  \langle |c_{n\ell_{j}}|^2 \rangle  =
\frac{1}{2j+1}\sum_{m_l} |c_{n \ell m_l}|^2 \sum_{m_j} |C^{jm_j}_{\ell m_l, s m_s}|^2 \,,   \label{Eq:Rel-from-non-rel}
\end{equation}
where $C^{jm_j}_{\ell m_l, s m_s}$ are the  Clebsch-Gordan coefficients with  $s=1/2$ and $\ell=3$. For an isotropic environment, 
$|c_{n \ell m_l}|^2 = \langle |c_{n \ell}|^2 \rangle$,
so that Eq.~\eqref{Eq:Rel-from-non-rel} simplifies to $\langle |c_{n\ell_{j}}|^2 \rangle  = \langle |c_{n \ell}|^2 \rangle/2$. Thereby, $\langle |c_{5f_{7/2}}|^2 \rangle = \langle |c_{5f_{5/2}}|^2 \rangle = \langle |c_{5f}|^2 \rangle/2$.

We computed the required hyperfine matrix elements using our {\em ab initio} relativistic  atomic-structure code with random-phase-approximation and Brueckner-orbital many-body contributions~\cite{tan2023precision,CamRadKuz12}. The results are (in atomic units):  
$\langle 5f_{5/2}||T||5f_{5/2}\rangle = 0.29$,
$\langle 5f_{7/2}||T||5f_{7/2}\rangle = 0.17$, 
$\langle 5f_{7/2}||T||5f_{5/2}\rangle =-0.097$. This implies that with a $\sim 20\%$ accuracy we may retain only the $j_h=j_d=5/2$ contribution to the rate~\eqref{eq:IC_rate_Relativistic}.

Then the IC rate can be expressed in terms of the hyperfine structure constant $A_{5f_{5/2}}$ for the \thor$^{3+}$  $5f_{5/2}$ ground state with the nucleus in its ground state by employing the relation~\cite{Joh07}
\begin{align}
A_{5f_{5/2}} =\frac{2}{5} \sqrt{\frac{2}{105}} \mu_g 
\langle 5f_{5/2}||T||5f_{5/2}\rangle \,,
\label{Eq:HFconsts}%
\end{align}
leading to an expected IC rate of
\begin{align}
\label{eq:IC_rate_finito}
\Gamma_{\mathrm{IC}}& \approx
 \frac{2\pi}{\hbar} \xi 
\rho_{5f}\left(  \varepsilon_{h_{r}}\right) 
\left( A_{5f_{5/2}} \right)^2 \,.
\end{align}
Here  $\xi$ is the dimensionless geometric factor
\begin{align}
\xi = \frac{125}{32}\frac{\langle|c_{5f_{5/2}}|^2\rangle}{\langle |c_{5f}|^2 \rangle} \frac{\left\vert 
\langle g||\mu||e\rangle\right\vert ^{2}}{ \mu_g^2 } \,,
\end{align}
with $\mu_g = 0.360(7) \mu_N$ being the  magnetic moment of the ground nuclear state~\cite{SafSafRad2013-Th3plus} and 
$\langle g||\mu||e\rangle=0.84\mu_N$ where $\mu_N$ is the nuclear magneton (see SI). 
As an estimate, in an isotropic environment, 
${\langle|c_{5f_{5/2}}|^2\rangle}/{\langle |c_{5f}|^2 \rangle} = 1/2$, resulting in $\xi \approx 11$. 
Using the measured value~\cite{Campbell2011} of the $5f_{5/2}$ hyperfine constant is $A_{5f_{5/2}} =  82.2(6) \, \mathrm{MHz}$, the IC decay rate in practical units is:
\begin{equation}
    \Gamma_{\rm IC} \approx 1.2 \times 10^4 \frac{\rho_{5f}\left(  \varepsilon_{h_{r}}\right)}{\text{states}/\text{eV}}  \,\, \text{s}^{-1} \,,
 \label{Eq:IC_rate_practical}   
\end{equation}
where $\rho_{5f}\left(  \varepsilon_{h_{r}} \right)$ is the Th impurity $f$-state PDOS (expressed as the number of states per eV) at the position of the resonant hole  $\varepsilon_{h_{r}} = \varepsilon_{d} - \omega_\mathrm{nuc} - E_R$. 

Now we return to the Ref.~\cite{Tiedau2024} statement ``since the nuclei are present in the Th$^{4+}$ charge state, we do not expect a significant contribution of bound internal conversion to the decay rate.''  It would have been correct only for vanishingly small values of 
$\rho_{5f}\left(  \varepsilon_{h_{r}} \right)$ in the IC rate~\eqref{Eq:IC_rate_practical}. This is not the case. Indeed, the PDOS order of magnitude can be estimated  by dividing PDOS integrated over the valence band by a typical few eV width of the valence band {(our DFT yields the width of $\approx 3.5 \, \mathrm{eV}$ for \lisaf{} and  $\approx 3 \, \mathrm{eV}$ for \ce{CaF2}, see Figs.~\ref{fig:valence band Th PDOS} and S2}). The integrated PDOS has a meaning of the effective number of Th $f$-electrons contributing to the valence band. Qualitatively, in ionic hosts, the valence band electrons scatter off of $\mathrm{Th}^{4+}$ core, so the integrated PDOS $\sim 1$ and we expect $\rho_{5f}\sim 0.1 \,\text{state/eV}$ in agreement with our DFT calculations for both \thor:\ce{LiSrAlF6} and \thor:\ce{CaF2}. This estimate leads to $\Gamma_{\rm IC} \sim 10^3 \,\text{s}^{-1}$. Thereby, if the IC process is energetically allowed in solid-state hosts, it  quenches the \thor{} isomer on a millisecond timescale, much faster than the isomer's measured~\cite{Tiedau2024,Elwell2024} $\sim 10^3 \, \mathrm{s}$ radiative lifetime in solid-state hosts. 

While we used the DFT calculations as an important qualitative guide in our derivation of the IC rate, predicting if the IC channel is open critically depends on the reliability of computing the defect state energies $\varepsilon_d$. This may require going beyond DFT methods that typically  underestimate excited state energies. A recent CASPT2 study~\cite{Nalikowski2025-Th-CaF2-Molcas}, using a cluster model with point charge and \textit{ab initio} model potential embedding, found defect state energies of $\sim 11\, \mathrm{eV}$ for Th:\ce{CaF2}, i.e. above $\omega_\mathrm{nuc}$, for relevant thorium environments. {Such an embedded cluster approach, in particular, includes particle-hole interactions~\cite{Dresselhaus2018} affecting the defect state energies.} {Further discussion of the effect of particle-hole interactions on the IC rate can be found in the SI.}

To summarize, if the IC process is energetically allowed, it rapidly quenches the excited nucleus. The recent experiments~\cite{Tiedau2024,Elwell2024,Zhang2024-Th229Comb} relied on observing a nuclear decay on a much longer time-scale. We conclude, that in these experiments, the IC was avoided and the doping sites contributing to the observed fluorescence had either 
\begin{itemize}
    \item[(i)] a sufficiently {\em high} energy of the defect state, $\varepsilon_d > \omega_\mathrm{nuc}$, or
    \item[(ii)]  a sufficiently {\em 
    low} energy of the defect state, so that the resonant hole energy lies in the bandgap between valence bands (e.g.,  below the valence band minimum, $\varepsilon_d - \varepsilon_\mathrm{min} > \omega_\mathrm{nuc}$.)
\end{itemize}
The condition (ii) would have been satisfied if the resonant hole energy ended up in the -3.6 eV -- -4 eV gap or below -4.5 eV on the Th PDOS plot, Fig.~\ref{fig:valence band Th PDOS}. {These conditions elucidate crucial requirements in choosing the solid-state hosts for the \thor{} doping in nuclear clock applications.}

{
While this paper was under review, an experiment~\cite{Terhune2024-photoquenching} on photo-quenching of the \thor{} isomer has been reported in \lisaf{}. The suggested quenching mechanism~\cite{Terhune2024-photoquenching} involves an IC process that proceeds by exciting the off-resonant laser populated electronic defect state into the conduction band. The estimate~\eqref{Eq:IC_rate_practical} immediately applies with $\rho_{5f}$ having the meaning of the {\em 
conduction} band $5f$-state PDOS. Our estimated  IC rate of $10^3 \,\mathrm{s}^{-1}$ agrees with the Ref.~\cite{Terhune2024-photoquenching} extracted value of $2\pi \times 10^2\,\mathrm{s}^{-1}$.
}

{
Finally, we point out that the IC process can determine the  critical doping density of Th above which the crystal can no longer serve as a nuclear clock host. Indeed, any interaction between the localized defect state electrons lowers the defect energy as a consequence of the degenerate perturbation theory. Various defect-defect interactions grow stronger with the reduced distances between the dopants~\cite{Mahan2000-book}. Thereby, for sufficiently high \thor{} doping concentration an IC channel may become open making the host unusable for the nuclear clock applications.}

\section*{Acknowledgements}
The authors thank Yafis  Barlas for valuable discussions.
This work was supported by NSF awards PHYS-2013011 and PHY-2207546, and ARO award W911NF-11-1-0369.
This work used Bridges-2 at Pittsburgh Supercomputing Center through allocation PHY230110 from the ACCESS program, which is supported by NSF grants \#2138259, \#2138286, \#2138307, \#2137603, and \#2138296.

\section{Supporting Information}

\subsection{Density functional theory methods}
DFT calculations were performed with VASP~\cite{RN12}, version 6.3, using the PAW~\cite{RN14} method with a plane-wave cutoff of 500 eV and a spin-restricted formalism.
The lowest-energy geometry for a Th atom in a supercell of \lisaf{} was determined by screening structures from a previous study of thorium-doped \ce{LiCaAlF6}~\cite{RN460}.
1625 structures of Th:\ce{LiCaAlF6}, in $2 \times 2 \times 1$ supercells of the host material, were converted to Th:\lisaf, using optimized lattice parameters for pure \lisaf, and reoptimized.
Energy corrections to account for differences in stochiometry were computed using optimized structure energies of binary and ternary fluorides of Li, Sr, and Al.
The lowest energy defect geometry was found to be Th replacing Sr with two interstitial F atoms for charge balancing.
This structure was expanded to a $3 \times 3 \times 2$ supercell of \lisaf{} and reoptimized.
All electronic structure properties were computed for this $3 \times 3 \times 2$ supercell.
The PBE~\cite{RN13} functional was used for all structural optimizations, and the modified Becke-Johnson (MBJ)~\cite{RN489,RN490} functional was used for electronic properties.
Optimizations of $2 \times 2 \times 1$ supercells were done with a 4-4-4 $k$-point grid, and electronic structure calculations on the $3 \times 3 \times 2$ supercell used a 4-4-2 $k$-point grid unless otherwise specified.
This large grid was previously determined to be necessary for accurate density of states computations~\cite{Elwell2024}.
Integration of the PDOS was done with the trapezium integration function in numpy.

Calculations on Th:\ce{CaF2} were done using a $3 \times 3 \times 3$ supercell of the conventional unit cell of \ce{CaF2}.
The thorium atom was put on a Ca site and two fluorine atoms were added to compensate the charge in the 90\degree configuration.
This has been found to be the lowest energy geometry in previous investigations and our own calculations.\cite{RN611}

The phonon analysis required optimization of an excited electronic state of Th:\ce{LiSrAlF6}.
To emphasize the local nature of the non-periodic Th defect environment, excited state optimizations were done with only the thorium atom and either 9 or 12 fluorine atoms free to move.
The excited state optimizations were done in a $2 \times 2 \times 1$ supercell of Th:\ce{LiSrAlF6} with the same local Th environment as in the larger supercells used in this study, using the $\Gamma$ point only.
The first electronic excited state was reached by fixing the occupancies of the Kohn-Sham orbitals such that an electron was promoted from the highest occupied orbital of the ground electronic state to the lowest unoccupied orbital of the ground state.
The first excited state, rather than an excited state on resonance with the nuclear transition, was used to ease excited state SCF convergence.
Both excited states involve F-Th electron transfer so they can be expected to have similar lattice relaxation parameters.
These calculations used an unrestricted Kohn-Sham wavefunction ansatz.
This is in the spirit of $\Delta$-SCF and was done in VASP using the ISMEAR settings.
Relaxation energies for Huang-Rhys theory analysis were computed by recalculating the ground electronic state at the optimized geometry of the excited state, and then taking the energy difference between that and the equilibrium geometry for the ground electronic state.

\subsection{Additional computational figures}
\begin{figure}[h!]
    \centering
    \includegraphics[width=\linewidth]{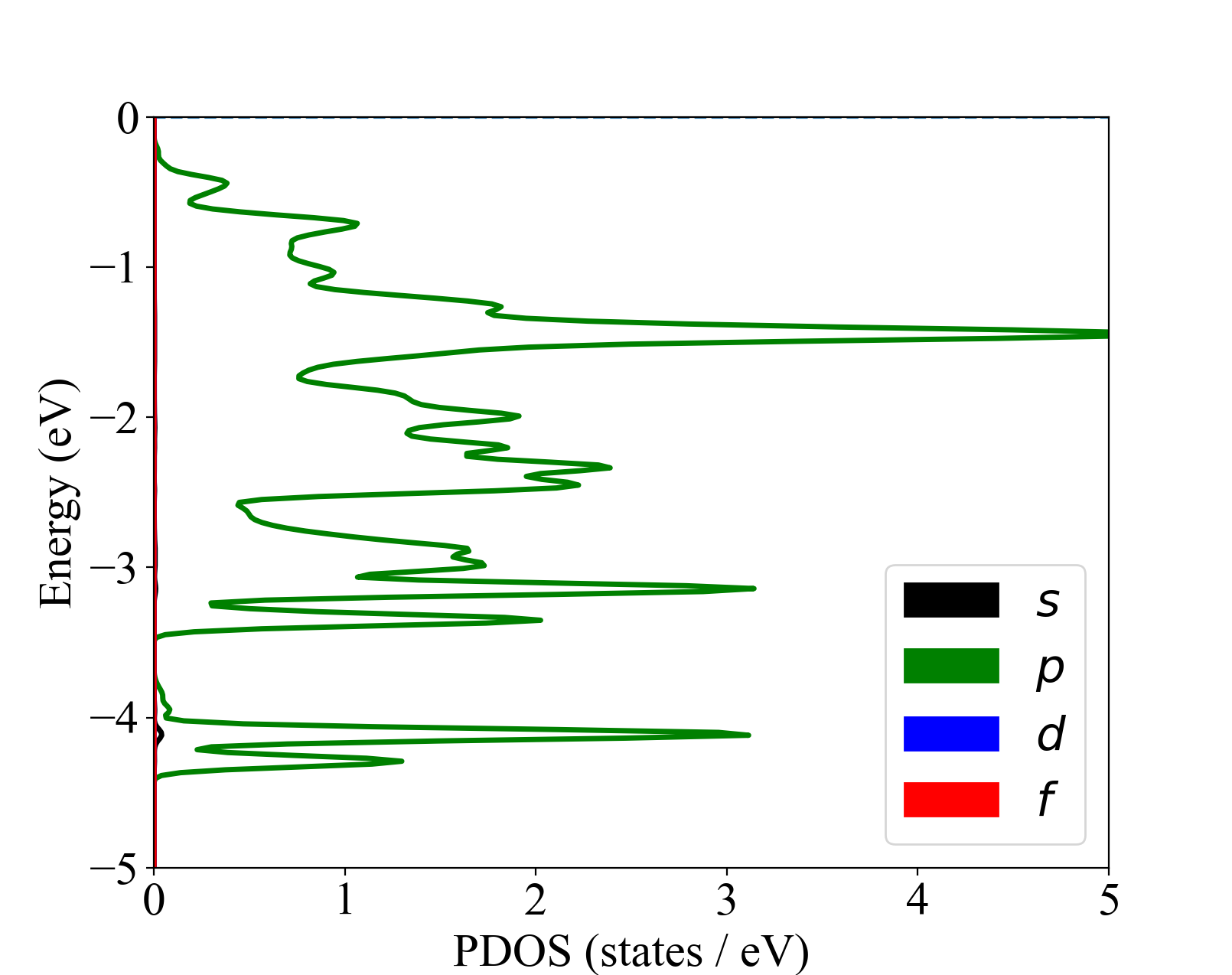}
    \caption{Fluorine projected density of states in the valence band for a F atom in Th:\ce{LiSrAlF6}.}
    \label{fig:ThLiSAF F PDOS}
\end{figure}

\begin{figure}[h!]
    \centering
    \includegraphics[width=\linewidth]{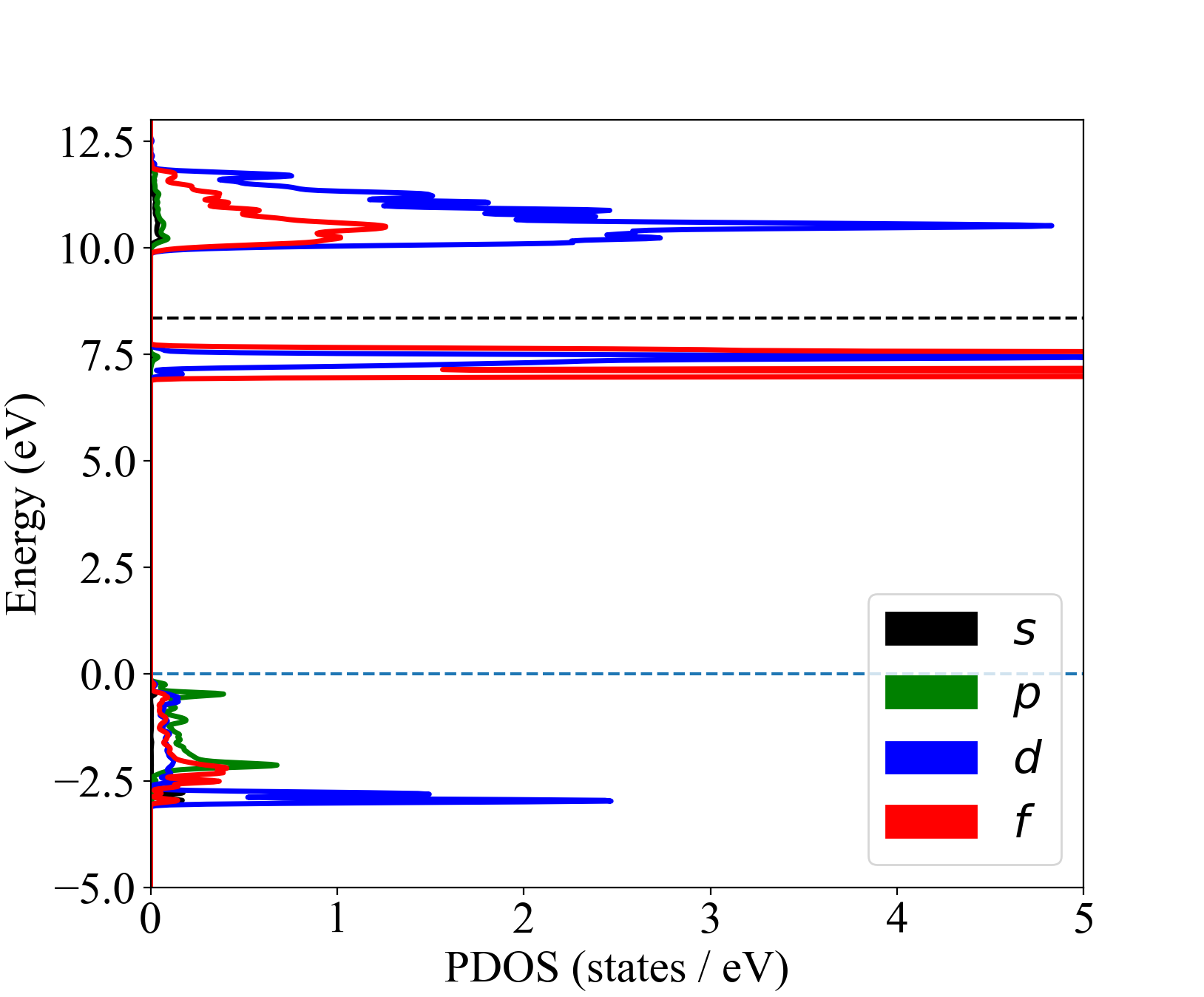}
    \caption{Thorium projected density of states in the for Th:\ce{CaF2}. The black dashed line marks the nuclear transition energy relative to the top of the valence band.}
    \label{fig:ThCaF2 Th PDOS}
\end{figure}

\begin{figure}[h!]
    \centering
    \includegraphics[width=\linewidth]{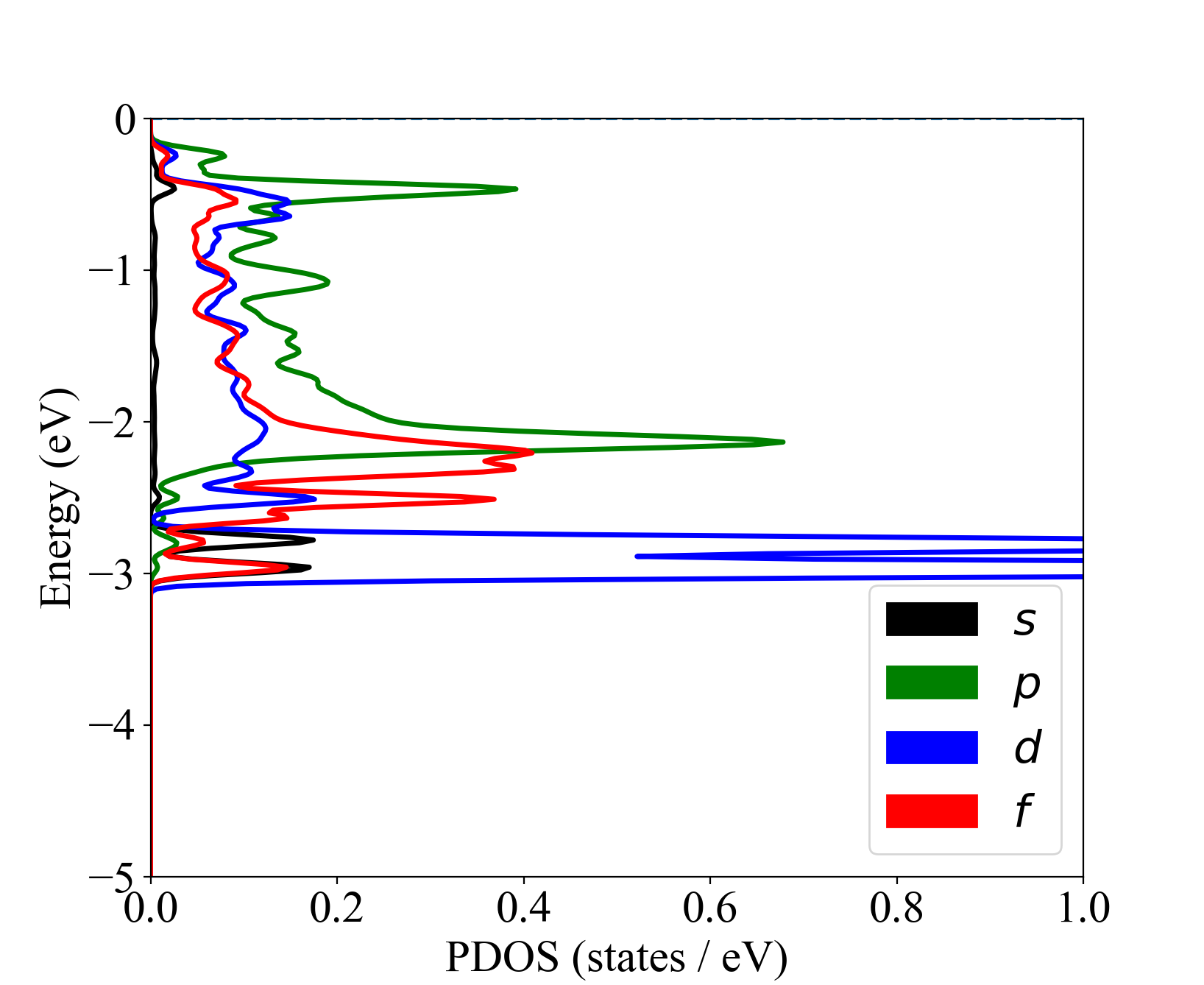}
    \caption{Thorium projected density of states in the valence band for Th:\ce{CaF2}.}
    \label{fig:ThCaF2 Th VB PDOS}
\end{figure}

\begin{figure}[h!]
    \centering
    \includegraphics[width=\linewidth]{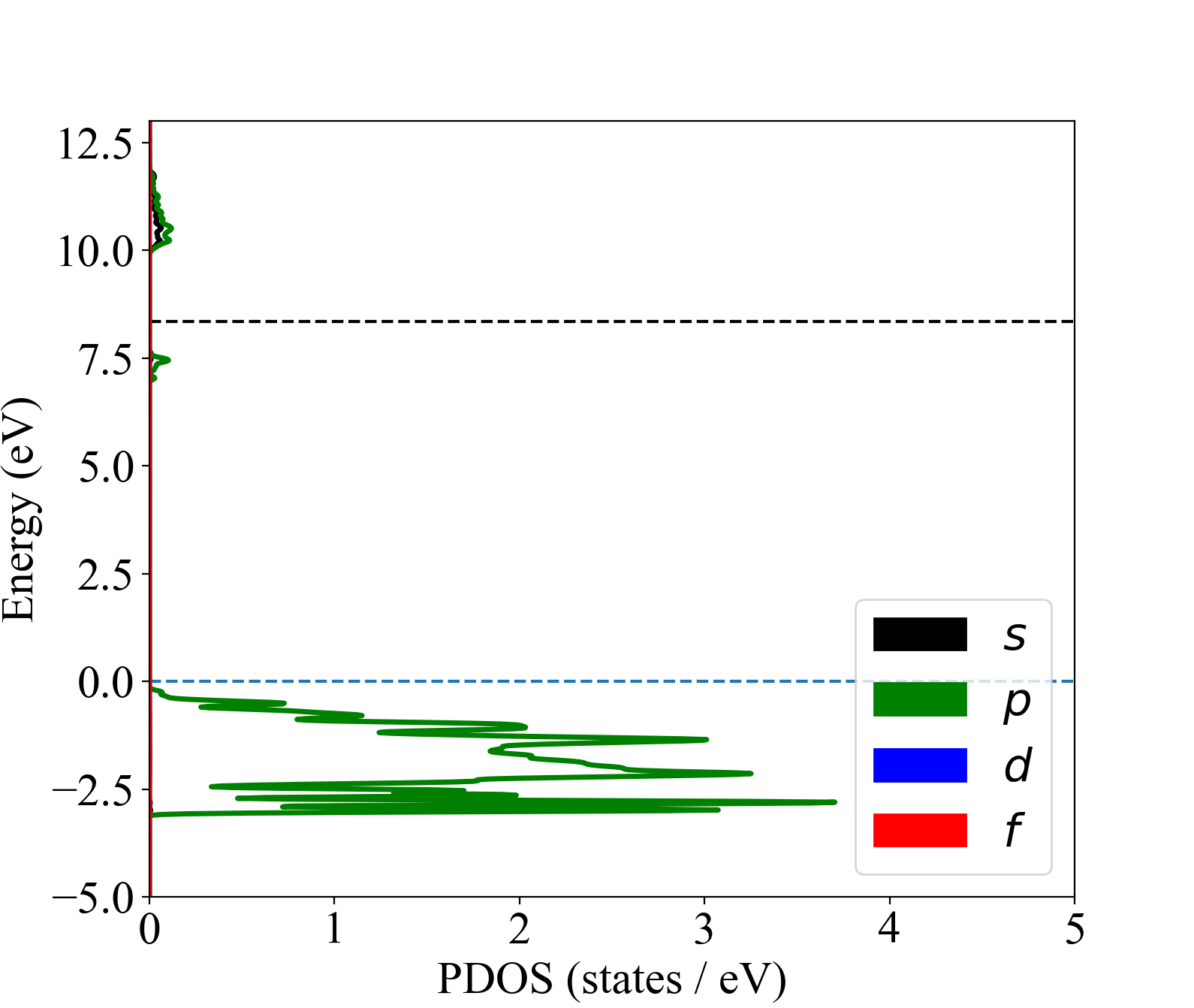}
    \caption{Fluorine projected density of states for a F atom in Th:\ce{CaF2}. The black dashed line marks the nuclear transition energy relative to the top of the valence band.}
    \label{fig:enter-label}
\end{figure}

See Figs.~\ref{fig:ThLiSAF F PDOS}, \ref{fig:ThCaF2 Th PDOS},
\ref{fig:ThCaF2 Th VB PDOS}, and \ref{fig:enter-label}.

\subsection{Hyperfine interaction}
{
General {\em ab initio} relativistic treatment of hyperfine interaction can be found in Ref.~\cite{Joh07}.
Magnetic-dipole hyperfine interaction  can be represented as a scalar product of the nuclear M1 operator $\boldsymbol{\mu}$  and the electronic  operator 
$\boldsymbol{T}$
\begin{equation*}
V = \boldsymbol{\mu}\cdot \boldsymbol{T} = 
\sum_{\lambda=-1}^{+1} (-1)^\lambda \boldsymbol{\mu}_{-\lambda} \boldsymbol{T}_\lambda \,,
\end{equation*}
Both operators are irreducible tensors of rank 1 (vectors) and 
we expanded the scalar product as a sum over their spherical components.
For a many-electron system, the electronic tensor $
\boldsymbol{T}_{\lambda} =\sum_{i}\boldsymbol{t}_{\lambda}\left(  \mathbf{r}_{i}\right)$,
with one-electron operators  
\begin{equation*}
t_{\lambda}\left(  \mathbf{r}\right)  = 
-\frac{i}{r^{2}}\sqrt{2}\boldsymbol{ \alpha}\cdot\mathbf{C}_{1,\lambda
}^{(0)}\left(  \mathbf{\hat{r}}\right) \,.
\end{equation*}
Here $\boldsymbol{ \alpha}$ is the Dirac matrix and $\mathbf{C}_{1,\lambda 
}^{(0)}\left(  \mathbf{\hat{r}}\right)$ are the normalized vector spherical harmonics and  atomic units $\hbar= m_e = |e| \equiv 
 1$ were used.
}

\subsection{Basis-independence of the total IC rate}
{A referee has asked the following question:
``In Eq. (8), the authors performed a summation of possible
decay channels for magnetic quantum numbers independently. However, I
guess that if the defect states are not energetically well separated,
the final defect state will be a superposition of them. If so, the
interference between the decay channel can strongly affect the decay
rate. Why the interference is not considered?''}

{
The rate is the probability of making a transition per unit time. If there are multiple final states $f$, probabilities for mutually-exclusive events add, and one sums over probabilities per unit time (rates) for all the decay channels, see e.g. textbook~\cite{GriffithsQM-book}: 
$$
 \Gamma = \sum_f \gamma_{i\rightarrow f} \,.
$$
Suppose the final states  form a degenerate manifold 
$\{|f\rangle\}$. We can linearly transform (rotate) the $\{|f\rangle\}$ basis  into a different basis $\{|f'\rangle\}$.
Then, as the referee implies, the decay would formally populate different linear superpositions as viewed in the rotated basis. However, the {\em total} rate $\Gamma$, the quantity of interest to us, is independent of the basis selection. Indeed, under a unitary transformation (basis rotation), $\sum_f |f\rangle  \langle f| = \sum_{f'} |f'\rangle  \langle f'|$. Then (c.f. 
Eq.~(8) of the main text, with $V$ being the hyperfine interaction), 
\begin{align*}
 \Gamma \propto & \sum_f |\langle i | V | f \rangle|^2  = 
 \sum_f \langle i | V | f \rangle \langle f | V | i \rangle 
 = \\ &
 \langle i | V \left(\sum_f | f \rangle \langle f |\right)  V | i \rangle = 
 \langle i | V \left(\sum_{f'} | f' \rangle \langle f' |\right)  V | i \rangle = \\&
 \sum_{f'} |\langle i | V | f' \rangle|^2 \, .
\end{align*}
This proves  that the expression for {\em total} rate is 
basis-independent. In particular, it means that the potential interference terms, i.e. cross terms involving different final substates, vanish identically.
}

\subsection{Effects of particle-hole interaction}
{
We estimate that the particle-hole interactions (beyond shifting the defect energies as discussed in the main text) do not substantially modify the IC rate. 
}

{
We derived the corrected IC rate in the second-order of perturbation theory, 
$\Gamma'_\text{IC} = \Gamma_\text{IC} \times ( 1 + |\Delta|^2)$, where 
$\Delta = i \frac{\alpha}{4\pi \varepsilon_\infty}\frac{c}{v_h} \log(\frac{v_h \tau_h}{a_d}) $,
with the high-frequency dielectric constant $\varepsilon_\infty \approx 2$, group velocity $v_h \approx 100 \, \text{km/s}$, typical hole  lifetime $\tau_h \sim \text{ns}$ and defect size $a_d \sim \text{Bohr radius}$. We find that  $|\Delta| \gg 1$, indicating the failure of the  perturbative treatment. Nevertheless, this exercise is valuable, as it demonstrates that the second-order correction {\em increases} the IC rate, which only strengthens the conclusion of our work: IC quenches the isomer on timescales much faster than the isomer's radiative lifetime.
}

{
Given the breakdown of the perturbative treatment, a nonperturbative (all-orders) treatment of the particle–hole interaction is required.  Such an approach can be implemented numerically via GW and Bethe–Salpeter equation (BSE) Green's-function techniques.  However, these calculations remain prohibitively expensive for the systems of interest, even on the most powerful servers at our disposal.  Nevertheless, the outcomes of a fully nonperturbative treatment can still be inferred through qualitative arguments.
}

{
Indeed, in our approach, we transfer the many-body solid-state physics to the computation of Th $f$-states Projected Density of States (PDOS) $\rho_{5f}$, while the hyperfine interaction matrix elements are evaluated using high-quality relativistic many-body methods of atomic structure, which include self-energy and random-phase-approximation diagrams. Then the major effect of defect-hole  interaction (beyond shifting the defect energies) is in modifying the PDOS.  Qualitatively, the PDOS order of magnitude can be estimated by dividing PDOS integrated over the valence band by a typical few eV width of the valence band. The integrated PDOS has a meaning of the effective number of Th $f$-electrons contributing to the valence band. In our hosts, the valence band electrons scatter off the $\mathrm{Th}^{4+}$ core, so the integrated PDOS $\sim 1$ and we expect $\rho_{5f}\sim 0.1 \,\text{state/eV}$ regardless of the defect-hole interaction effects. This is consistent with the PDOS values used in our manuscript for estimating the IC rate.
}

{
Based on this qualitative argument, we conclude that the particle-hole interactions do not substantially modify our estimated IC rate.
}

\subsection{Nuclear Excitation by Electron Transition}
{
Nuclear Excitation by Electron Transition (NEET) is the time‐reverse analogue of internal conversion  (IC), the focus of  our paper. In NEET, a particle-hole electronic transition feeds energy into the nucleus, promoting it from its ground state to an excited (isomeric) state.
The NEET and IC rates are related by the principle of detailed balance.
}

{
The NEET process requires a pre-formed particle-hole pair. For a \thor-doped crystal, the electronic defect state must be pre-populated together with a residual hole in the valence band. The energy of such a pair must be resonant with the nuclear transition. At room temperatures, probabilities of thermally populating such NEET pairs are exceedingly low as $\hbar \omega_\text{nuc}/k_B T \sim 3 \times 10^2$.
}

{
The NEET resonant particle-hole pairs can form as a result of the IC process. However, the IC creates a hole state $h_r$ that is hot, i.e. with energy below the valence band maximum, see Fig.~1 of the main text. Such hot holes ``float'' to the top of the valence band on {\em picosecond} timescales typical for non-radiative relaxation of hot holes~\cite{Woerner1993}
due to vibronic couplings.
This relaxation mechanism rapidly breaks the NEET resonance condition. In addition, the radiative particle-hole lifetimes in a solid are on the order of a nanosecond. Both of these mechanisms are much faster than the millisecond time scale of the NEET process, rapidly depleting the IC-spawned resonance pairs. 
Thereby, we do not expect any sizable correction to the computed IC rate. This discussion suggests that with a high degree of probability the IC process terminates  in a photon emitted by radiative recombination of the defect state electron with a hole at the top of the valence band. This observation enables the spectroscopic determination of the defect state energy (see also Ref.~\cite{Elwell2024}). %
}

To summarize, the NEET process does not affect the manuscript conclusions.

\subsection{Lifetimes and the off-diagonal M1 nuclear matrix element}
The new measurements~\cite{Tiedau2024,Elwell2024} of the energy and radiative lifetime of the 229-Th isomeric state imply an estimate for the M1 matrix element for 229-Th somewhat different from its current value of $\langle g||\mu||e\rangle=1.2\mu_N$~\cite{Dykhne1998,Porsev2010}. While there is very good agreement between Refs.~\cite{Tiedau2024} and \cite{Elwell2024} on the isomeric state energy of $\omega\approx8.355$ eV, their values for the isomeric lifetime, $\tau=2510\pm 72$ s and $\tau=1860\pm79$ s differ significantly. To get an estimate for $\langle g||\mu||e\rangle$, we seek to combine these two values for $\tau$.

One may adopt a simple approach for a weighted average and chi-squared-rescaled uncertainty where
\begin{subequations}
    \begin{align}
        \tau&=\frac{w_1\tau_1+w_2\tau_2}{w_1+w_2}\,,\\
        \chi^2&=w_1(\tau_1-\tau)^2+w_2(\tau_2-\tau)^2\,,\\
        \sigma^2 &= \frac{\chi^2}{w_1+w_2}\,,
    \end{align}
\end{subequations}
with $\tau_1=2510$ s, $\tau_2=1860$ s, $w_1=1/\sigma_1^2=1/72$ s${}^{-2}$, and $w_2=1/\sigma_2^2=1/79$ s${}^{-2}$. This however, assumes similar confidence in the measurements. 

Since the measurement of $\tau$ depends on knowledge of the crystal index of refraction, which is much better characterized for CaF${}_2$~\cite{Tiedau2024} than for LiSrAlF${}_6$~\cite{Elwell2024}, we rescale the weight $w_1$ by a factor $k>1$. Here, we pick $k=2$, thus obtaining

\begin{equation}
    \tau=2319\pm296\,\,{\rm s}\,.
\end{equation}

Using $\tau=3/(\omega^3\langle g||\mu||e\rangle^2)$, one thus finds
\begin{equation}
    \langle g||\mu||e\rangle=(0.84\pm0.11)\mu_N\,.
\end{equation}


\begin{thebibliography}{36}%
\makeatletter
\providecommand \@ifxundefined [1]{%
 \@ifx{#1\undefined}
}%
\providecommand \@ifnum [1]{%
 \ifnum #1\expandafter \@firstoftwo
 \else \expandafter \@secondoftwo
 \fi
}%
\providecommand \@ifx [1]{%
 \ifx #1\expandafter \@firstoftwo
 \else \expandafter \@secondoftwo
 \fi
}%
\providecommand \natexlab [1]{#1}%
\providecommand \enquote  [1]{``#1''}%
\providecommand \bibnamefont  [1]{#1}%
\providecommand \bibfnamefont [1]{#1}%
\providecommand \citenamefont [1]{#1}%
\providecommand \href@noop [0]{\@secondoftwo}%
\providecommand \href [0]{\begingroup \@sanitize@url \@href}%
\providecommand \@href[1]{\@@startlink{#1}\@@href}%
\providecommand \@@href[1]{\endgroup#1\@@endlink}%
\providecommand \@sanitize@url [0]{\catcode `\\12\catcode `\$12\catcode
  `\&12\catcode `\#12\catcode `\^12\catcode `\_12\catcode `\%12\relax}%
\providecommand \@@startlink[1]{}%
\providecommand \@@endlink[0]{}%
\providecommand \url  [0]{\begingroup\@sanitize@url \@url }%
\providecommand \@url [1]{\endgroup\@href {#1}{\urlprefix }}%
\providecommand \urlprefix  [0]{URL }%
\providecommand \Eprint [0]{\href }%
\providecommand \doibase [0]{https://doi.org/}%
\providecommand \selectlanguage [0]{\@gobble}%
\providecommand \bibinfo  [0]{\@secondoftwo}%
\providecommand \bibfield  [0]{\@secondoftwo}%
\providecommand \translation [1]{[#1]}%
\providecommand \BibitemOpen [0]{}%
\providecommand \bibitemStop [0]{}%
\providecommand \bibitemNoStop [0]{.\EOS\space}%
\providecommand \EOS [0]{\spacefactor3000\relax}%
\providecommand \BibitemShut  [1]{\csname bibitem#1\endcsname}%
\let\auto@bib@innerbib\@empty
\bibitem [{\citenamefont {Rellergert}\ \emph {et~al.}(2010)\citenamefont
  {Rellergert}, \citenamefont {DeMille}, \citenamefont {Greco}, \citenamefont
  {Hehlen}, \citenamefont {Torgerson},\ and\ \citenamefont
  {Hudson}}]{Rellergert2010}%
  \BibitemOpen
  \bibfield  {author} {\bibinfo {author} {\bibfnamefont {W.~G.}\ \bibnamefont
  {Rellergert}}, \bibinfo {author} {\bibfnamefont {D.}~\bibnamefont {DeMille}},
  \bibinfo {author} {\bibfnamefont {R.~R.}\ \bibnamefont {Greco}}, \bibinfo
  {author} {\bibfnamefont {M.~P.}\ \bibnamefont {Hehlen}}, \bibinfo {author}
  {\bibfnamefont {J.~R.}\ \bibnamefont {Torgerson}},\ and\ \bibinfo {author}
  {\bibfnamefont {E.~R.}\ \bibnamefont {Hudson}},\ }\href
  {https://doi.org/10.1103/PhysRevLett.104.200802} {\bibfield  {journal}
  {\bibinfo  {journal} {Phys. Rev. Lett.}\ }\textbf {\bibinfo {volume} {104}},\
  \bibinfo {pages} {200802} (\bibinfo {year} {2010})}\BibitemShut {NoStop}%
\bibitem [{\citenamefont {Tiedau}\ \emph {et~al.}(2024)\citenamefont {Tiedau},
  \citenamefont {Okhapkin}, \citenamefont {Zhang}, \citenamefont {Thielking},
  \citenamefont {Zitzer}, \citenamefont {Peik}, \citenamefont {Schaden},
  \citenamefont {Pronebner}, \citenamefont {Morawetz}, \citenamefont {Toscani
  De~Col} \emph {et~al.}}]{Tiedau2024}%
  \BibitemOpen
  \bibfield  {author} {\bibinfo {author} {\bibfnamefont {J.}~\bibnamefont
  {Tiedau}}, \bibinfo {author} {\bibfnamefont {M.~V.}\ \bibnamefont
  {Okhapkin}}, \bibinfo {author} {\bibfnamefont {K.}~\bibnamefont {Zhang}},
  \bibinfo {author} {\bibfnamefont {J.}~\bibnamefont {Thielking}}, \bibinfo
  {author} {\bibfnamefont {G.}~\bibnamefont {Zitzer}}, \bibinfo {author}
  {\bibfnamefont {E.}~\bibnamefont {Peik}}, \bibinfo {author} {\bibfnamefont
  {F.}~\bibnamefont {Schaden}}, \bibinfo {author} {\bibfnamefont
  {T.}~\bibnamefont {Pronebner}}, \bibinfo {author} {\bibfnamefont
  {I.}~\bibnamefont {Morawetz}}, \bibinfo {author} {\bibfnamefont
  {L.}~\bibnamefont {Toscani De~Col}}, \emph {et~al.},\ }\href
  {https://journals.aps.org/prl/accepted/2c07aYbeC981d47c171619f5604116053962ac79a}
  {\bibfield  {journal} {\bibinfo  {journal} {Phys. Rev. Lett.}\ }\textbf
  {\bibinfo {volume} {132}},\ \bibinfo {pages} {182501} (\bibinfo {year}
  {2024})}\BibitemShut {NoStop}%
\bibitem [{\citenamefont {Elwell}\ \emph {et~al.}(2024)\citenamefont {Elwell},
  \citenamefont {Schneider}, \citenamefont {Jeet}, \citenamefont {Terhune},
  \citenamefont {Morgan}, \citenamefont {Alexandrova}, \citenamefont {{Tran
  Tan}}, \citenamefont {Derevianko},\ and\ \citenamefont
  {Hudson}}]{Elwell2024}%
  \BibitemOpen
  \bibfield  {author} {\bibinfo {author} {\bibfnamefont {R.}~\bibnamefont
  {Elwell}}, \bibinfo {author} {\bibfnamefont {C.}~\bibnamefont {Schneider}},
  \bibinfo {author} {\bibfnamefont {J.}~\bibnamefont {Jeet}}, \bibinfo {author}
  {\bibfnamefont {J.~E.~S.}\ \bibnamefont {Terhune}}, \bibinfo {author}
  {\bibfnamefont {H.~W.~T.}\ \bibnamefont {Morgan}}, \bibinfo {author}
  {\bibfnamefont {A.~N.}\ \bibnamefont {Alexandrova}}, \bibinfo {author}
  {\bibfnamefont {H.~B.}\ \bibnamefont {{Tran Tan}}}, \bibinfo {author}
  {\bibfnamefont {A.}~\bibnamefont {Derevianko}},\ and\ \bibinfo {author}
  {\bibfnamefont {E.~R.}\ \bibnamefont {Hudson}},\ }\href@noop {} {\bibfield
  {journal} {\bibinfo  {journal} {Phys. Rev. Lett.}\ }\textbf {\bibinfo
  {volume} {133}},\ \bibinfo {pages} {013201} (\bibinfo {year}
  {2024})}\BibitemShut {NoStop}%
\bibitem [{\citenamefont {Zhang}\ \emph
  {et~al.}(2024{\natexlab{a}})\citenamefont {Zhang}, \citenamefont {von~der
  Wense}, \citenamefont {Doyle}, \citenamefont {Higgins}, \citenamefont {Ooi},
  \citenamefont {Friebel}, \citenamefont {Ye}, \citenamefont {Elwell},
  \citenamefont {Terhune}, \citenamefont {Morgan}, \citenamefont {Alexandrova},
  \citenamefont {{Tran Tan}}, \citenamefont {Derevianko},\ and\ \citenamefont
  {Hudson}}]{zhang2024thf}%
  \BibitemOpen
  \bibfield  {author} {\bibinfo {author} {\bibfnamefont {C.}~\bibnamefont
  {Zhang}}, \bibinfo {author} {\bibfnamefont {L.}~\bibnamefont {von~der
  Wense}}, \bibinfo {author} {\bibfnamefont {J.~F.}\ \bibnamefont {Doyle}},
  \bibinfo {author} {\bibfnamefont {J.~S.}\ \bibnamefont {Higgins}}, \bibinfo
  {author} {\bibfnamefont {T.}~\bibnamefont {Ooi}}, \bibinfo {author}
  {\bibfnamefont {H.~U.}\ \bibnamefont {Friebel}}, \bibinfo {author}
  {\bibfnamefont {J.}~\bibnamefont {Ye}}, \bibinfo {author} {\bibfnamefont
  {R.}~\bibnamefont {Elwell}}, \bibinfo {author} {\bibfnamefont {J.~E.~S.}\
  \bibnamefont {Terhune}}, \bibinfo {author} {\bibfnamefont {H.~W.~T.}\
  \bibnamefont {Morgan}}, \bibinfo {author} {\bibfnamefont {A.~N.}\
  \bibnamefont {Alexandrova}}, \bibinfo {author} {\bibfnamefont {H.~B.}\
  \bibnamefont {{Tran Tan}}}, \bibinfo {author} {\bibfnamefont
  {A.}~\bibnamefont {Derevianko}},\ and\ \bibinfo {author} {\bibfnamefont
  {E.~R.}\ \bibnamefont {Hudson}},\ }\href
  {https://doi.org/10.1038/s41586-024-08256-5} {\bibfield  {journal} {\bibinfo
  {journal} {Nature}\ }\textbf {\bibinfo {volume} {636}},\ \bibinfo {pages}
  {603} (\bibinfo {year} {2024}{\natexlab{a}})}\BibitemShut {NoStop}%
\bibitem [{\citenamefont {Zhang}\ \emph
  {et~al.}(2024{\natexlab{b}})\citenamefont {Zhang}, \citenamefont {Ooi},
  \citenamefont {Higgins}, \citenamefont {Doyle}, \citenamefont {von~der
  Wense}, \citenamefont {Beeks}, \citenamefont {Leitner}, \citenamefont
  {Kazakov}, \citenamefont {Li}, \citenamefont {Thirolf}, \citenamefont
  {Schumm},\ and\ \citenamefont {Ye}}]{Zhang2024-Th229Comb}%
  \BibitemOpen
  \bibfield  {author} {\bibinfo {author} {\bibfnamefont {C.}~\bibnamefont
  {Zhang}}, \bibinfo {author} {\bibfnamefont {T.}~\bibnamefont {Ooi}}, \bibinfo
  {author} {\bibfnamefont {J.~S.}\ \bibnamefont {Higgins}}, \bibinfo {author}
  {\bibfnamefont {J.~F.}\ \bibnamefont {Doyle}}, \bibinfo {author}
  {\bibfnamefont {L.}~\bibnamefont {von~der Wense}}, \bibinfo {author}
  {\bibfnamefont {K.}~\bibnamefont {Beeks}}, \bibinfo {author} {\bibfnamefont
  {A.}~\bibnamefont {Leitner}}, \bibinfo {author} {\bibfnamefont {G.~A.}\
  \bibnamefont {Kazakov}}, \bibinfo {author} {\bibfnamefont {P.}~\bibnamefont
  {Li}}, \bibinfo {author} {\bibfnamefont {P.~G.}\ \bibnamefont {Thirolf}},
  \bibinfo {author} {\bibfnamefont {T.}~\bibnamefont {Schumm}},\ and\ \bibinfo
  {author} {\bibfnamefont {J.}~\bibnamefont {Ye}},\ }\href
  {https://doi.org/10.1038/s41586-024-07839-6} {\bibfield  {journal} {\bibinfo
  {journal} {Nature}\ }\textbf {\bibinfo {volume} {633}},\ \bibinfo {pages}
  {63} (\bibinfo {year} {2024}{\natexlab{b}})},\ \Eprint
  {https://arxiv.org/abs/2406.18719} {arXiv:2406.18719} \BibitemShut {NoStop}%
\bibitem [{\citenamefont {Karpeshin}\ and\ \citenamefont
  {Trzhaskovskaya}(2007)}]{Karpeshin2007}%
  \BibitemOpen
  \bibfield  {author} {\bibinfo {author} {\bibfnamefont {F.~F.}\ \bibnamefont
  {Karpeshin}}\ and\ \bibinfo {author} {\bibfnamefont {M.~B.}\ \bibnamefont
  {Trzhaskovskaya}},\ }\href {https://doi.org/10.1103/PhysRevC.76.054313}
  {\bibfield  {journal} {\bibinfo  {journal} {Phys. Rev. C}\ }\textbf {\bibinfo
  {volume} {76}},\ \bibinfo {pages} {1} (\bibinfo {year} {2007})}\BibitemShut
  {NoStop}%
\bibitem [{\citenamefont {Tkalya}\ \emph
  {et~al.}(2015{\natexlab{a}})\citenamefont {Tkalya}, \citenamefont
  {Schneider}, \citenamefont {Jeet},\ and\ \citenamefont
  {Hudson}}]{Tkalya2015}%
  \BibitemOpen
  \bibfield  {author} {\bibinfo {author} {\bibfnamefont {E.~V.}\ \bibnamefont
  {Tkalya}}, \bibinfo {author} {\bibfnamefont {C.}~\bibnamefont {Schneider}},
  \bibinfo {author} {\bibfnamefont {J.}~\bibnamefont {Jeet}},\ and\ \bibinfo
  {author} {\bibfnamefont {E.~R.}\ \bibnamefont {Hudson}},\ }\href
  {https://doi.org/10.1103/PhysRevC.92.054324} {\bibfield  {journal} {\bibinfo
  {journal} {Phys. Rev. C}\ }\textbf {\bibinfo {volume} {92}},\ \bibinfo
  {pages} {054324} (\bibinfo {year} {2015}{\natexlab{a}})}\BibitemShut
  {NoStop}%
\bibitem [{\citenamefont {Bilous}\ \emph {et~al.}(2017)\citenamefont {Bilous},
  \citenamefont {Kazakov}, \citenamefont {Moore}, \citenamefont {Schumm},\ and\
  \citenamefont {P{\'{a}}lffy}}]{Bilous2017}%
  \BibitemOpen
  \bibfield  {author} {\bibinfo {author} {\bibfnamefont {P.~V.}\ \bibnamefont
  {Bilous}}, \bibinfo {author} {\bibfnamefont {G.~A.}\ \bibnamefont {Kazakov}},
  \bibinfo {author} {\bibfnamefont {I.~D.}\ \bibnamefont {Moore}}, \bibinfo
  {author} {\bibfnamefont {T.}~\bibnamefont {Schumm}},\ and\ \bibinfo {author}
  {\bibfnamefont {A.}~\bibnamefont {P{\'{a}}lffy}},\ }\href
  {https://doi.org/10.1103/PhysRevA.95.032503} {\bibfield  {journal} {\bibinfo
  {journal} {Physical Review A}\ }\textbf {\bibinfo {volume} {95}},\ \bibinfo
  {pages} {1} (\bibinfo {year} {2017})}\BibitemShut {NoStop}%
\bibitem [{\citenamefont {Bilous}\ \emph {et~al.}(2018)\citenamefont {Bilous},
  \citenamefont {Minkov},\ and\ \citenamefont {P{\'{a}}lffy}}]{Bilous2018}%
  \BibitemOpen
  \bibfield  {author} {\bibinfo {author} {\bibfnamefont {P.~V.}\ \bibnamefont
  {Bilous}}, \bibinfo {author} {\bibfnamefont {N.}~\bibnamefont {Minkov}},\
  and\ \bibinfo {author} {\bibfnamefont {A.}~\bibnamefont {P{\'{a}}lffy}},\
  }\href {https://doi.org/10.1103/PhysRevC.97.044320} {\bibfield  {journal}
  {\bibinfo  {journal} {Phys. Rev. C}\ }\textbf {\bibinfo {volume} {97}},\
  \bibinfo {pages} {2} (\bibinfo {year} {2018})}\BibitemShut {NoStop}%
\bibitem [{\citenamefont {Tkalya}\ \emph
  {et~al.}(2015{\natexlab{b}})\citenamefont {Tkalya}, \citenamefont
  {Schneider}, \citenamefont {Jeet},\ and\ \citenamefont
  {Hudson}}]{TkaSchJee2015}%
  \BibitemOpen
  \bibfield  {author} {\bibinfo {author} {\bibfnamefont {E.~V.}\ \bibnamefont
  {Tkalya}}, \bibinfo {author} {\bibfnamefont {C.}~\bibnamefont {Schneider}},
  \bibinfo {author} {\bibfnamefont {J.}~\bibnamefont {Jeet}},\ and\ \bibinfo
  {author} {\bibfnamefont {E.~R.}\ \bibnamefont {Hudson}},\ }\href
  {https://doi.org/10.1103/PhysRevC.92.054324} {\bibfield  {journal} {\bibinfo
  {journal} {Phys. Rev. C}\ }\textbf {\bibinfo {volume} {92}},\ \bibinfo
  {pages} {054324} (\bibinfo {year} {2015}{\natexlab{b}})},\ \Eprint
  {https://arxiv.org/abs/1509.09101} {arXiv:1509.09101} \BibitemShut {NoStop}%
\bibitem [{\citenamefont {Pineda}\ \emph {et~al.}(2024)\citenamefont {Pineda},
  \citenamefont {Chhetri}, \citenamefont {Bara}, \citenamefont {Elskens},
  \citenamefont {Casci}, \citenamefont {Alexandrova}, \citenamefont {Au},
  \citenamefont {Athanasakis-Kaklamanakis}, \citenamefont {Bartokos},
  \citenamefont {Beeks}, \citenamefont {Bernerd}, \citenamefont {Claessens},
  \citenamefont {Chrysalidis}, \citenamefont {Cocolios}, \citenamefont
  {Correia}, \citenamefont {{De Witte}}, \citenamefont {Elwell}, \citenamefont
  {Ferrer}, \citenamefont {Heinke}, \citenamefont {Hudson}, \citenamefont
  {Ivandikov}, \citenamefont {Kudryavtsev}, \citenamefont {K{\"{o}}ster},
  \citenamefont {Kraemer}, \citenamefont {Laatiaoui}, \citenamefont {Lica},
  \citenamefont {Merckling}, \citenamefont {Morawetz}, \citenamefont {Morgan},
  \citenamefont {Moritz}, \citenamefont {Pereira}, \citenamefont {Raeder},
  \citenamefont {Rothe}, \citenamefont {Schaden}, \citenamefont {Scharl},
  \citenamefont {Schumm}, \citenamefont {Stegemann}, \citenamefont {Terhune},
  \citenamefont {Thirolf}, \citenamefont {Tunhuma}, \citenamefont {Bergh},
  \citenamefont {{Van Duppen}}, \citenamefont {Vantomme}, \citenamefont
  {Wahl},\ and\ \citenamefont {Yue}}]{Pineda2024}%
  \BibitemOpen
  \bibfield  {author} {\bibinfo {author} {\bibfnamefont {S.~V.}\ \bibnamefont
  {Pineda}}, \bibinfo {author} {\bibfnamefont {P.}~\bibnamefont {Chhetri}},
  \bibinfo {author} {\bibfnamefont {S.}~\bibnamefont {Bara}}, \bibinfo {author}
  {\bibfnamefont {Y.}~\bibnamefont {Elskens}}, \bibinfo {author} {\bibfnamefont
  {S.}~\bibnamefont {Casci}}, \bibinfo {author} {\bibfnamefont {A.~N.}\
  \bibnamefont {Alexandrova}}, \bibinfo {author} {\bibfnamefont
  {M.}~\bibnamefont {Au}}, \bibinfo {author} {\bibfnamefont {M.}~\bibnamefont
  {Athanasakis-Kaklamanakis}}, \bibinfo {author} {\bibfnamefont
  {M.}~\bibnamefont {Bartokos}}, \bibinfo {author} {\bibfnamefont
  {K.}~\bibnamefont {Beeks}}, \bibinfo {author} {\bibfnamefont
  {C.}~\bibnamefont {Bernerd}}, \bibinfo {author} {\bibfnamefont
  {A.}~\bibnamefont {Claessens}}, \bibinfo {author} {\bibfnamefont
  {K.}~\bibnamefont {Chrysalidis}}, \bibinfo {author} {\bibfnamefont {T.~E.}\
  \bibnamefont {Cocolios}}, \bibinfo {author} {\bibfnamefont {J.~G.}\
  \bibnamefont {Correia}}, \bibinfo {author} {\bibfnamefont {H.}~\bibnamefont
  {{De Witte}}}, \bibinfo {author} {\bibfnamefont {R.}~\bibnamefont {Elwell}},
  \bibinfo {author} {\bibfnamefont {R.}~\bibnamefont {Ferrer}}, \bibinfo
  {author} {\bibfnamefont {R.}~\bibnamefont {Heinke}}, \bibinfo {author}
  {\bibfnamefont {E.~R.}\ \bibnamefont {Hudson}}, \bibinfo {author}
  {\bibfnamefont {F.}~\bibnamefont {Ivandikov}}, \bibinfo {author}
  {\bibfnamefont {Y.}~\bibnamefont {Kudryavtsev}}, \bibinfo {author}
  {\bibfnamefont {U.}~\bibnamefont {K{\"{o}}ster}}, \bibinfo {author}
  {\bibfnamefont {S.}~\bibnamefont {Kraemer}}, \bibinfo {author} {\bibfnamefont
  {M.}~\bibnamefont {Laatiaoui}}, \bibinfo {author} {\bibfnamefont
  {R.}~\bibnamefont {Lica}}, \bibinfo {author} {\bibfnamefont {C.}~\bibnamefont
  {Merckling}}, \bibinfo {author} {\bibfnamefont {I.}~\bibnamefont {Morawetz}},
  \bibinfo {author} {\bibfnamefont {H.~W.~T.}\ \bibnamefont {Morgan}}, \bibinfo
  {author} {\bibfnamefont {D.}~\bibnamefont {Moritz}}, \bibinfo {author}
  {\bibfnamefont {L.~M.~C.}\ \bibnamefont {Pereira}}, \bibinfo {author}
  {\bibfnamefont {S.}~\bibnamefont {Raeder}}, \bibinfo {author} {\bibfnamefont
  {S.}~\bibnamefont {Rothe}}, \bibinfo {author} {\bibfnamefont
  {F.}~\bibnamefont {Schaden}}, \bibinfo {author} {\bibfnamefont
  {K.}~\bibnamefont {Scharl}}, \bibinfo {author} {\bibfnamefont
  {T.}~\bibnamefont {Schumm}}, \bibinfo {author} {\bibfnamefont
  {S.}~\bibnamefont {Stegemann}}, \bibinfo {author} {\bibfnamefont
  {J.}~\bibnamefont {Terhune}}, \bibinfo {author} {\bibfnamefont {P.~G.}\
  \bibnamefont {Thirolf}}, \bibinfo {author} {\bibfnamefont {S.~M.}\
  \bibnamefont {Tunhuma}}, \bibinfo {author} {\bibfnamefont {P.~V.~D.}\
  \bibnamefont {Bergh}}, \bibinfo {author} {\bibfnamefont {P.}~\bibnamefont
  {{Van Duppen}}}, \bibinfo {author} {\bibfnamefont {A.}~\bibnamefont
  {Vantomme}}, \bibinfo {author} {\bibfnamefont {U.}~\bibnamefont {Wahl}},\
  and\ \bibinfo {author} {\bibfnamefont {Z.}~\bibnamefont {Yue}},\ }\href
  {http://arxiv.org/abs/2408.12309} {\ ,\ \bibinfo {pages} {1} (\bibinfo {year}
  {2024})},\ \Eprint {https://arxiv.org/abs/2408.12309} {arXiv:2408.12309}
  \BibitemShut {NoStop}%
\bibitem [{\citenamefont {Fetter}\ and\ \citenamefont
  {Walecka}(1971)}]{FetWal71}%
  \BibitemOpen
  \bibfield  {author} {\bibinfo {author} {\bibfnamefont {A.~L.}\ \bibnamefont
  {Fetter}}\ and\ \bibinfo {author} {\bibfnamefont {J.~D.}\ \bibnamefont
  {Walecka}},\ }\href@noop {} {\emph {\bibinfo {title} {{Quantum Theory of
  Many-particle Systems}}}}\ (\bibinfo  {publisher} {McGraw-Hill},\ \bibinfo
  {address} {New York},\ \bibinfo {year} {1971})\BibitemShut {NoStop}%
\bibitem [{\citenamefont {Cohen-Tannoudji}\ \emph
  {et~al.}(1998{\natexlab{a}})\citenamefont {Cohen-Tannoudji}, \citenamefont
  {Diu}, \citenamefont {Laloe},\ and\ \citenamefont {Merzbacher}}]{CohTan98}%
  \BibitemOpen
  \bibfield  {author} {\bibinfo {author} {\bibfnamefont {C.}~\bibnamefont
  {Cohen-Tannoudji}}, \bibinfo {author} {\bibfnamefont {B.}~\bibnamefont
  {Diu}}, \bibinfo {author} {\bibfnamefont {F.}~\bibnamefont {Laloe}},\ and\
  \bibinfo {author} {\bibfnamefont {E.}~\bibnamefont {Merzbacher}},\
  }\href@noop {} {\emph {\bibinfo {title} {{Quantum Mechanics}}}},\ \bibinfo
  {edition} {3rd}\ ed.,\ Vol.\ \bibinfo {volume} {I and II}\ (\bibinfo
  {publisher} {Wiley, John and Sons},\ \bibinfo {year} {1998})\BibitemShut
  {NoStop}%
\bibitem [{\citenamefont {Cohen-Tannoudji}\ \emph
  {et~al.}(1998{\natexlab{b}})\citenamefont {Cohen-Tannoudji}, \citenamefont
  {Dupont-Roc},\ and\ \citenamefont {Grynberg}}]{cohen1998atom}%
  \BibitemOpen
  \bibfield  {author} {\bibinfo {author} {\bibfnamefont {C.}~\bibnamefont
  {Cohen-Tannoudji}}, \bibinfo {author} {\bibfnamefont {J.}~\bibnamefont
  {Dupont-Roc}},\ and\ \bibinfo {author} {\bibfnamefont {G.}~\bibnamefont
  {Grynberg}},\ }\href
  {https://onlinelibrary.wiley.com/doi/book/10.1002/9783527617197} {\emph
  {\bibinfo {title} {Atom-photon interactions: basic processes and
  applications}}}\ (\bibinfo  {publisher} {John Wiley \& Sons},\ \bibinfo
  {year} {1998})\BibitemShut {NoStop}%
\bibitem [{\citenamefont {{VASP Software GmbH}}(2023)}]{VASP}%
  \BibitemOpen
  \bibfield  {author} {\bibinfo {author} {\bibnamefont {{VASP Software
  GmbH}}},\ }\href {https://www.vasp.at} {\bibinfo {title} {{Vienna Ab initio
  Simulation Package (VASP)}}} (\bibinfo {year} {2023})\BibitemShut {NoStop}%
\bibitem [{\citenamefont {Andrews}\ \emph {et~al.}(2013)\citenamefont
  {Andrews}, \citenamefont {Thanthiriwatte}, \citenamefont {Wang},\ and\
  \citenamefont {Dixon}}]{Andrews2013}%
  \BibitemOpen
  \bibfield  {author} {\bibinfo {author} {\bibfnamefont {L.}~\bibnamefont
  {Andrews}}, \bibinfo {author} {\bibfnamefont {K.~S.}\ \bibnamefont
  {Thanthiriwatte}}, \bibinfo {author} {\bibfnamefont {X.}~\bibnamefont
  {Wang}},\ and\ \bibinfo {author} {\bibfnamefont {D.~A.}\ \bibnamefont
  {Dixon}},\ }\href {https://doi.org/10.1021/ic401107w} {\bibfield  {journal}
  {\bibinfo  {journal} {Inorganic Chemistry}\ }\textbf {\bibinfo {volume}
  {52}},\ \bibinfo {pages} {8228} (\bibinfo {year} {2013})}\BibitemShut
  {NoStop}%
\bibitem [{\citenamefont {Dessovic}\ \emph {et~al.}(2014)\citenamefont
  {Dessovic}, \citenamefont {Mohn}, \citenamefont {Jackson}, \citenamefont
  {Winkler}, \citenamefont {Schreitl}, \citenamefont {Kazakov},\ and\
  \citenamefont {Schumm}}]{RN611}%
  \BibitemOpen
  \bibfield  {author} {\bibinfo {author} {\bibfnamefont {P.}~\bibnamefont
  {Dessovic}}, \bibinfo {author} {\bibfnamefont {P.}~\bibnamefont {Mohn}},
  \bibinfo {author} {\bibfnamefont {R.}~\bibnamefont {Jackson}}, \bibinfo
  {author} {\bibfnamefont {G.}~\bibnamefont {Winkler}}, \bibinfo {author}
  {\bibfnamefont {M.}~\bibnamefont {Schreitl}}, \bibinfo {author}
  {\bibfnamefont {G.}~\bibnamefont {Kazakov}},\ and\ \bibinfo {author}
  {\bibfnamefont {T.}~\bibnamefont {Schumm}},\ }\bibfield  {journal} {\bibinfo
  {journal} {Journal of Physics - Condensed Matter}\ }\textbf {\bibinfo
  {volume} {26}},\ \href {https://doi.org/10.1088/0953-8984/26/10/105402}
  {10.1088/0953-8984/26/10/105402} (\bibinfo {year} {2014})\BibitemShut
  {NoStop}%
\bibitem [{\citenamefont {{Tran Tan}}\ and\ \citenamefont
  {Derevianko}(2023)}]{tan2023precision}%
  \BibitemOpen
  \bibfield  {author} {\bibinfo {author} {\bibfnamefont {H.~B.}\ \bibnamefont
  {{Tran Tan}}}\ and\ \bibinfo {author} {\bibfnamefont {A.}~\bibnamefont
  {Derevianko}},\ }\href {https://doi.org/10.1103/PhysRevA.107.042809}
  {\bibfield  {journal} {\bibinfo  {journal} {Phys. Rev. A}\ }\textbf {\bibinfo
  {volume} {107}},\ \bibinfo {pages} {042809} (\bibinfo {year}
  {2023})}\BibitemShut {NoStop}%
\bibitem [{\citenamefont {Campbell}\ \emph {et~al.}(2012)\citenamefont
  {Campbell}, \citenamefont {Radnaev}, \citenamefont {Kuzmich}, \citenamefont
  {Dzuba}, \citenamefont {Flambaum},\ and\ \citenamefont
  {Derevianko}}]{CamRadKuz12}%
  \BibitemOpen
  \bibfield  {author} {\bibinfo {author} {\bibfnamefont {C.~J.}\ \bibnamefont
  {Campbell}}, \bibinfo {author} {\bibfnamefont {A.~G.}\ \bibnamefont
  {Radnaev}}, \bibinfo {author} {\bibfnamefont {A.}~\bibnamefont {Kuzmich}},
  \bibinfo {author} {\bibfnamefont {V.~A.}\ \bibnamefont {Dzuba}}, \bibinfo
  {author} {\bibfnamefont {V.~V.}\ \bibnamefont {Flambaum}},\ and\ \bibinfo
  {author} {\bibfnamefont {A.}~\bibnamefont {Derevianko}},\ }\href
  {https://doi.org/10.1103/PhysRevLett.108.120802} {\bibfield  {journal}
  {\bibinfo  {journal} {Phys. Rev. Lett.}\ }\textbf {\bibinfo {volume} {108}},\
  \bibinfo {pages} {120802} (\bibinfo {year} {2012})}\BibitemShut {NoStop}%
\bibitem [{\citenamefont {Johnson}(2007)}]{Joh07}%
  \BibitemOpen
  \bibfield  {author} {\bibinfo {author} {\bibfnamefont {W.~R.}\ \bibnamefont
  {Johnson}},\ }\href
  {https://link.springer.com/book/10.1007/978-3-540-68013-0} {\emph {\bibinfo
  {title} {{Atomic Structure Theory: Lectures on Atomic Physics}}}}\ (\bibinfo
  {publisher} {Springer},\ \bibinfo {address} {New York, NY},\ \bibinfo {year}
  {2007})\BibitemShut {NoStop}%
\bibitem [{\citenamefont {Safronova}\ \emph {et~al.}(2013)\citenamefont
  {Safronova}, \citenamefont {Safronova}, \citenamefont {Radnaev},
  \citenamefont {Campbell},\ and\ \citenamefont
  {Kuzmich}}]{SafSafRad2013-Th3plus}%
  \BibitemOpen
  \bibfield  {author} {\bibinfo {author} {\bibfnamefont {M.~S.}\ \bibnamefont
  {Safronova}}, \bibinfo {author} {\bibfnamefont {U.~I.}\ \bibnamefont
  {Safronova}}, \bibinfo {author} {\bibfnamefont {A.~G.}\ \bibnamefont
  {Radnaev}}, \bibinfo {author} {\bibfnamefont {C.~J.}\ \bibnamefont
  {Campbell}},\ and\ \bibinfo {author} {\bibfnamefont {A.}~\bibnamefont
  {Kuzmich}},\ }\href {https://doi.org/10.1103/physreva.88.060501} {\bibfield
  {journal} {\bibinfo  {journal} {Physical Review A}\ }\textbf {\bibinfo
  {volume} {88}},\ \bibinfo {pages} {6} (\bibinfo {year} {2013})}\BibitemShut
  {NoStop}%
\bibitem [{\citenamefont {Campbell}\ \emph {et~al.}(2011)\citenamefont
  {Campbell}, \citenamefont {Radnaev},\ and\ \citenamefont
  {Kuzmich}}]{Campbell2011}%
  \BibitemOpen
  \bibfield  {author} {\bibinfo {author} {\bibfnamefont {C.}~\bibnamefont
  {Campbell}}, \bibinfo {author} {\bibfnamefont {A.}~\bibnamefont {Radnaev}},\
  and\ \bibinfo {author} {\bibfnamefont {A.}~\bibnamefont {Kuzmich}},\ }\href
  {https://doi.org/10.1103/PhysRevLett.106.223001} {\bibfield  {journal}
  {\bibinfo  {journal} {Phys. Rev. Lett.}\ }\textbf {\bibinfo {volume} {106}},\
  \bibinfo {pages} {223001} (\bibinfo {year} {2011})}\BibitemShut {NoStop}%
\bibitem [{\citenamefont {Nalikowski}\ \emph {et~al.}(2025)\citenamefont
  {Nalikowski}, \citenamefont {Veryazov}, \citenamefont {Beeks}, \citenamefont
  {Schumm},\ and\ \citenamefont
  {Kro{\'{s}}nicki}}]{Nalikowski2025-Th-CaF2-Molcas}%
  \BibitemOpen
  \bibfield  {author} {\bibinfo {author} {\bibfnamefont {K.}~\bibnamefont
  {Nalikowski}}, \bibinfo {author} {\bibfnamefont {V.}~\bibnamefont
  {Veryazov}}, \bibinfo {author} {\bibfnamefont {K.}~\bibnamefont {Beeks}},
  \bibinfo {author} {\bibfnamefont {T.}~\bibnamefont {Schumm}},\ and\ \bibinfo
  {author} {\bibfnamefont {M.}~\bibnamefont {Kro{\'{s}}nicki}},\ }\href
  {https://doi.org/10.1103/PhysRevB.111.115103} {\bibfield  {journal} {\bibinfo
   {journal} {Phys. Rev. B}\ }\textbf {\bibinfo {volume} {111}},\ \bibinfo
  {pages} {115103} (\bibinfo {year} {2025})},\ \Eprint
  {https://arxiv.org/abs/2410.00230} {arXiv:2410.00230} \BibitemShut {NoStop}%
\bibitem [{\citenamefont {Dresselhaus}\ \emph {et~al.}(2018)\citenamefont
  {Dresselhaus}, \citenamefont {Dresselhaus}, \citenamefont {Cronin},\ and\
  \citenamefont {{Gomes Souza Filho}}}]{Dresselhaus2018}%
  \BibitemOpen
  \bibfield  {author} {\bibinfo {author} {\bibfnamefont {M.}~\bibnamefont
  {Dresselhaus}}, \bibinfo {author} {\bibfnamefont {G.}~\bibnamefont
  {Dresselhaus}}, \bibinfo {author} {\bibfnamefont {S.~B.}\ \bibnamefont
  {Cronin}},\ and\ \bibinfo {author} {\bibfnamefont {A.}~\bibnamefont {{Gomes
  Souza Filho}}},\ }in\ \href {https://doi.org/10.1007/978-3-662-55922-2_20}
  {\emph {\bibinfo {booktitle} {Solid State Properties}}}\ (\bibinfo {address}
  {Berlin, Heidelberg},\ \bibinfo {year} {2018})\ Chap.~\bibinfo {chapter}
  {20}, pp.\ \bibinfo {pages} {411--441}\BibitemShut {NoStop}%
\bibitem [{\citenamefont {Terhune}\ \emph {et~al.}(2024)\citenamefont
  {Terhune}, \citenamefont {Elwell}, \citenamefont {Tan}, \citenamefont
  {Perera}, \citenamefont {Morgan}, \citenamefont {Alexandrova}, \citenamefont
  {Derevianko},\ and\ \citenamefont {Hudson}}]{Terhune2024-photoquenching}%
  \BibitemOpen
  \bibfield  {author} {\bibinfo {author} {\bibfnamefont {J.~E.~S.}\
  \bibnamefont {Terhune}}, \bibinfo {author} {\bibfnamefont {R.}~\bibnamefont
  {Elwell}}, \bibinfo {author} {\bibfnamefont {H.~B.~T.}\ \bibnamefont {Tan}},
  \bibinfo {author} {\bibfnamefont {U.~C.}\ \bibnamefont {Perera}}, \bibinfo
  {author} {\bibfnamefont {H.~W.~T.}\ \bibnamefont {Morgan}}, \bibinfo {author}
  {\bibfnamefont {A.~N.}\ \bibnamefont {Alexandrova}}, \bibinfo {author}
  {\bibfnamefont {A.}~\bibnamefont {Derevianko}},\ and\ \bibinfo {author}
  {\bibfnamefont {E.~R.}\ \bibnamefont {Hudson}},\ }\href
  {http://arxiv.org/abs/2412.08998} {\ ,\ \bibinfo {pages} {1} (\bibinfo {year}
  {2024})},\ \Eprint {https://arxiv.org/abs/2412.08998} {arXiv:2412.08998}
  \BibitemShut {NoStop}%
\bibitem [{\citenamefont {Mahan}(2000)}]{Mahan2000-book}%
  \BibitemOpen
  \bibfield  {author} {\bibinfo {author} {\bibfnamefont {G.~D.}\ \bibnamefont
  {Mahan}},\ }\href@noop {} {\emph {\bibinfo {title} {{Many-Particle
  Physics}}}},\ \bibinfo {edition} {3rd}\ ed.\ (\bibinfo  {publisher} {Kluwer
  Academic/Plenum Publishers},\ \bibinfo {address} {New York},\ \bibinfo {year}
  {2000})\BibitemShut {NoStop}%
\bibitem [{\citenamefont {Kresse}\ and\ \citenamefont
  {Furthmuller}(1996)}]{RN12}%
  \BibitemOpen
  \bibfield  {author} {\bibinfo {author} {\bibfnamefont {G.}~\bibnamefont
  {Kresse}}\ and\ \bibinfo {author} {\bibfnamefont {J.}~\bibnamefont
  {Furthmuller}},\ }\href {https://doi.org/10.1103/PhysRevB.54.11169}
  {\bibfield  {journal} {\bibinfo  {journal} {Physical Review B}\ }\textbf
  {\bibinfo {volume} {54}},\ \bibinfo {pages} {11169} (\bibinfo {year}
  {1996})}\BibitemShut {NoStop}%
\bibitem [{\citenamefont {Blochl}(1994)}]{RN14}%
  \BibitemOpen
  \bibfield  {author} {\bibinfo {author} {\bibfnamefont {P.~E.}\ \bibnamefont
  {Blochl}},\ }\href {https://doi.org/10.1103/PhysRevB.50.17953} {\bibfield
  {journal} {\bibinfo  {journal} {Physical Review B}\ }\textbf {\bibinfo
  {volume} {50}},\ \bibinfo {pages} {17953} (\bibinfo {year}
  {1994})}\BibitemShut {NoStop}%
\bibitem [{\citenamefont {Pimon}\ \emph {et~al.}(2022)\citenamefont {Pimon},
  \citenamefont {Mohn},\ and\ \citenamefont {Schumm}}]{RN460}%
  \BibitemOpen
  \bibfield  {author} {\bibinfo {author} {\bibfnamefont {M.}~\bibnamefont
  {Pimon}}, \bibinfo {author} {\bibfnamefont {P.}~\bibnamefont {Mohn}},\ and\
  \bibinfo {author} {\bibfnamefont {T.}~\bibnamefont {Schumm}},\ }\bibfield
  {journal} {\bibinfo  {journal} {Advanced Theory and Simulations}\ }\textbf
  {\bibinfo {volume} {5}},\ \href {https://doi.org/10.1002/adts.202200185}
  {10.1002/adts.202200185} (\bibinfo {year} {2022})\BibitemShut {NoStop}%
\bibitem [{\citenamefont {Perdew}\ \emph {et~al.}(1996)\citenamefont {Perdew},
  \citenamefont {Burke},\ and\ \citenamefont {Ernzerhof}}]{RN13}%
  \BibitemOpen
  \bibfield  {author} {\bibinfo {author} {\bibfnamefont {J.~P.}\ \bibnamefont
  {Perdew}}, \bibinfo {author} {\bibfnamefont {K.}~\bibnamefont {Burke}},\ and\
  \bibinfo {author} {\bibfnamefont {M.}~\bibnamefont {Ernzerhof}},\ }\href
  {https://doi.org/10.1103/PhysRevLett.77.3865} {\bibfield  {journal} {\bibinfo
   {journal} {Physical Review Letters}\ }\textbf {\bibinfo {volume} {77}},\
  \bibinfo {pages} {3865} (\bibinfo {year} {1996})}\BibitemShut {NoStop}%
\bibitem [{\citenamefont {Tran}\ and\ \citenamefont {Blaha}(2009)}]{RN489}%
  \BibitemOpen
  \bibfield  {author} {\bibinfo {author} {\bibfnamefont {F.}~\bibnamefont
  {Tran}}\ and\ \bibinfo {author} {\bibfnamefont {P.}~\bibnamefont {Blaha}},\
  }\href {https://doi.org/10.1103/PhysRevLett.102.226401} {\bibfield  {journal}
  {\bibinfo  {journal} {Phys. Rev. Lett.}\ }\textbf {\bibinfo {volume} {102}},\
  \bibinfo {pages} {226401} (\bibinfo {year} {2009})}\BibitemShut {NoStop}%
\bibitem [{\citenamefont {Becke}\ and\ \citenamefont {Johnson}(2006)}]{RN490}%
  \BibitemOpen
  \bibfield  {author} {\bibinfo {author} {\bibfnamefont {A.~D.}\ \bibnamefont
  {Becke}}\ and\ \bibinfo {author} {\bibfnamefont {E.~R.}\ \bibnamefont
  {Johnson}},\ }\href {https://doi.org/10.1063/1.2213970} {\bibfield  {journal}
  {\bibinfo  {journal} {J. Chem. Phys.}\ }\textbf {\bibinfo {volume} {124}},\
  \bibinfo {pages} {221101} (\bibinfo {year} {2006})}\BibitemShut {NoStop}%
\bibitem [{\citenamefont {Griffiths}\ and\ \citenamefont
  {Schroeter}(2018)}]{GriffithsQM-book}%
  \BibitemOpen
  \bibfield  {author} {\bibinfo {author} {\bibfnamefont {D.~J.}\ \bibnamefont
  {Griffiths}}\ and\ \bibinfo {author} {\bibfnamefont {D.~F.}\ \bibnamefont
  {Schroeter}},\ }\href@noop {} {\emph {\bibinfo {title} {{Introduction to
  Quantum Mechanics}}}},\ \bibinfo {edition} {3rd}\ ed.\ (\bibinfo  {publisher}
  {Cambridge University Press},\ \bibinfo {address} {Cambridge, UK},\ \bibinfo
  {year} {2018})\BibitemShut {NoStop}%
\bibitem [{\citenamefont {Woerner}\ \emph {et~al.}(1992)\citenamefont
  {Woerner}, \citenamefont {Elsaesser},\ and\ \citenamefont
  {Kaiser}}]{Woerner1993}%
  \BibitemOpen
  \bibfield  {author} {\bibinfo {author} {\bibfnamefont {M.}~\bibnamefont
  {Woerner}}, \bibinfo {author} {\bibfnamefont {T.}~\bibnamefont {Elsaesser}},\
  and\ \bibinfo {author} {\bibfnamefont {W.}~\bibnamefont {Kaiser}},\ }\href
  {https://doi.org/10.1103/PhysRevB.45.8378} {\bibfield  {journal} {\bibinfo
  {journal} {Physical Review B}\ }\textbf {\bibinfo {volume} {45}},\ \bibinfo
  {pages} {8378} (\bibinfo {year} {1992})}\BibitemShut {NoStop}%
\bibitem [{\citenamefont {Dykhne}\ and\ \citenamefont
  {Tkalya}(1998)}]{Dykhne1998}%
  \BibitemOpen
  \bibfield  {author} {\bibinfo {author} {\bibfnamefont {A.~M.}\ \bibnamefont
  {Dykhne}}\ and\ \bibinfo {author} {\bibfnamefont {E.~V.}\ \bibnamefont
  {Tkalya}},\ }\href {http://dx.doi.org/10.1134/1.567659} {\bibfield  {journal}
  {\bibinfo  {journal} {JETP Lett.}\ }\textbf {\bibinfo {volume} {67}},\
  \bibinfo {pages} {251} (\bibinfo {year} {1998})},\ \bibinfo {note} {[Pis'ma
  Zh. \'{E}ksp. Teor. Fiz. 67, 233--238 (1998)]}\BibitemShut {NoStop}%
\bibitem [{\citenamefont {Porsev}\ and\ \citenamefont
  {Flambaum}(2010)}]{Porsev2010}%
  \BibitemOpen
  \bibfield  {author} {\bibinfo {author} {\bibfnamefont {S.~G.}\ \bibnamefont
  {Porsev}}\ and\ \bibinfo {author} {\bibfnamefont {V.~V.}\ \bibnamefont
  {Flambaum}},\ }\href {https://doi.org/10.1103/PhysRevA.81.032504} {\bibfield
  {journal} {\bibinfo  {journal} {Phys. Rev. A}\ }\textbf {\bibinfo {volume}
  {81}},\ \bibinfo {pages} {032504} (\bibinfo {year} {2010})}\BibitemShut
  {NoStop}%
\end{thebibliography}

%

\end{document}